\documentclass[fleqn,11pt]{article}
\usepackage{amscd,amsmath,amssymb,verbatim}
\usepackage{amsthm}
\usepackage[pdftex]{graphicx}
\usepackage{epstopdf}

\usepackage{authblk}

\usepackage{epstopdf}

\usepackage[T1]{fontenc}
\usepackage{ae,aecompl}
\usepackage{subfigure}
\usepackage{array}
\usepackage{mathrsfs}
\usepackage[nohead]{geometry}
\usepackage[singlespacing]{setspace}
\usepackage{rotating}
\usepackage{pdflscape}
\usepackage{multirow}
\usepackage{threeparttable}
\usepackage{enumerate}
\usepackage{subfigure}
\usepackage{float}
\usepackage{placeins}
\usepackage{color}
\usepackage[table,xcdraw]{xcolor}
\usepackage{colortbl}
\usepackage{accents}
\usepackage{tabularx}
\usepackage{booktabs}
\usepackage{natbib}

\usepackage{float}
\numberwithin{equation}{section}

\DeclareMathOperator{\Diag}{Diag}

\newcommand{\bs}{\boldsymbol}
\newcommand{\E}{\mathbb{E}}

\newcommand{\mf}{\mathbf}

\newcommand{\xvec}{\boldsymbol}

\DeclareMathOperator{\e}{\varepsilon}
\usepackage{hyperref}

\makeatletter
\newcommand{\distas}[1]{\mathbin{\overset{#1}{\kern\z@\sim}}}%
\newsavebox{\mybox}\newsavebox{\mysim}
\newcommand{\distras}[1]{%
  \savebox{\mybox}{\hbox{\kern3pt$\scriptstyle#1$\kern3pt}}%
  \savebox{\mysim}{\hbox{$\sim$}}%
  \mathbin{\overset{#1}{\kern\z@\resizebox{\wd\mybox}{\ht\mysim}{$\sim$}}}%
}
\makeatother

\newtheorem{algorithm}{Algorithm}

\renewcommand{\hat}[1]{\widehat{\text{$#1$}}}

\makeatletter
\newsavebox\myboxA
\newsavebox\myboxB
\newlength\mylenA

\renewcommand*\bar[2][0.5]{%
    \sbox{\myboxA}{$\m@th#2$}%
    \setbox\myboxB\null
    \ht\myboxB=\ht\myboxA%
    \dp\myboxB=\dp\myboxA%
    \wd\myboxB=#1\wd\myboxA
    \sbox\myboxB{$\m@th\overline{\copy\myboxB}$}
    \setlength\mylenA{\the\wd\myboxA}
    \addtolength\mylenA{-\the\wd\myboxB}%
    \ifdim\wd\myboxB<\wd\myboxA%
       \rlap{\hskip 0.5\mylenA\usebox\myboxB}{\usebox\myboxA}%
    \else
        \hskip -0.5\mylenA\rlap{\usebox\myboxA}{\hskip 0.5\mylenA\usebox\myboxB}%
    \fi}
\makeatother
\hoffset 
\voffset

\date{\today}
\title{A Dynamic Spatiotemporal and Network ARCH Model with Common Factors}

\author[1]{Osman Do\u{g}an}
\author[2]{Raffaele Mattera}
\author[3]{Philipp Otto\thanks{Corresponding author: philipp.otto@glasgow.ac.uk}}
\author[4]{Süleyman Ta\c{s}p\i nar}

\affil[1]{Istanbul Technical University, Department of Economics, Turkey}
\affil[2]{University of Campania ``Luigi Vanvitelli'', Department of Mathematics and Physics, Italy}
\affil[3]{University of Glasgow, School of Mathematics and Statistics, UK}
\affil[4]{Queens College, City University New York, USA}

\begin{document}
\maketitle

\singlespacing

\begin{abstract}
\noindent  We introduce a dynamic spatiotemporal volatility model that extends traditional approaches by incorporating spatial, temporal, and spatiotemporal spillover effects, along with volatility-specific observed and latent factors. The model offers a more general network interpretation, making it applicable for studying various types of network spillovers.  The primary innovation lies in incorporating volatility-specific latent factors into the dynamic spatiotemporal volatility model. Using Bayesian estimation via the Markov Chain Monte Carlo (MCMC) method, the model offers a robust framework for analyzing the spatial, temporal, and spatiotemporal effects of a log-squared outcome variable on its volatility. We recommend using the deviance information criterion (DIC) and a regularized Bayesian MCMC method to select the number of relevant factors in the model. The model’s flexibility is demonstrated through two applications: a spatiotemporal model applied to the U.S. housing market and another applied to financial stock market networks, both highlighting the model’s ability to capture varying degrees of interconnectedness. In both applications, we find strong spatial/network interactions with relatively stronger spillover effects in the stock market.
\end{abstract}

\vspace{7cm}
\noindent
JEL-Classification: C11, C23, C58.
\onehalfspacing

\section{Introduction}\label{intro}

\noindent Volatility modeling plays a crucial role in understanding the dynamics of economic, financial, and other complex systems, as it captures the uncertainty and the risk associated with the temporal evolution of an underlying outcome variable. The most common example of heteroskedastic time series can be found in the financial domain, albeit evidence of dynamic volatility has been found in macroeconomics \citep[e.g.][]{kiani2004business}, house prices \citep[e.g.][]{bollerslev2016daily,dougan2023bayesian}, and also in environmental time series \citep[e.g.][]{huang2011class,otto2023dynamic}. Traditional approaches model volatility dynamics through autoregressive processes \citep[e.g., the GARCH,][]{bollerslev1986generalized} that often rely on the information of volatility proxies and observable exogenous variables. In the multivariate setting, multivariate GARCH models are adopted to study the relationships between volatilities and correlations among a set of time series. These models are helpful under the assumption that the volatility of a time series is transmitted to another so that the information on the volatility of each time series is used to model the volatility of the others. However, these models are often not suitable for high-dimensional setups, as they usually occur in spatial (i.e., having a large number of spatial locations) or network settings (i.e., having a large number of nodes) because long time series are required for estimation. Instead, spatiotemporal and network ARCH models assume a certain structure in the (instantaneous) cross-sectional dimension (spatial or network dimension), which can be interpreted in the standard GARCH-type sense \citep[see][for a review of spatial and spatiotemporal volatility models]{otto2024spatial}.

The dependencies between time series can be significantly strengthened by incorporating information on spatial proximity.  Tobler's First Law of Geography -- ``everything is related to everything else, but near things are more related than distant things'' -- suggests that the volatility dynamics of time series associated with spatial units that are geographically close are likely to exhibit stronger interdependencies than those that are farther apart. This implies that spatial proximity amplifies the connections between time series, intensifying the impact of local shocks and spillovers on volatility patterns. Traditional multivariate GARCH models, such as the BEKK \citep{engle1995multivariate} or the DCC \citep{engle2002dynamic}, do not allow for testing the existence of this kind of spillovers between the multivariate time series under study, nor quantifying precisely its magnitude. For financial time series, these instantaneous interactions in the volatility result from contemporaneous decisions of actors on the financial market. In other words, agents observe the financial risk of all (or most) financial assets and instantaneously account for these risks in their own trading decisions. Similar findings have been found for house prices \citep[e.g., see][]{cohen2016spatial,otto2018spatiotemporal}. In the same manner, for environmental applications, environmental conditions affect the variation in all locations contemporaneously, with a decreasing magnitude for an increasing geographical distance \citep[e.g., see][]{rodeschini2024scenario,otto2023dynamic}. In the case of macroeconomics, for example, there is widely documented evidence of the existence of spillovers between neighboring economies in terms of many indicators \citep{benos2015proximity,dell2013spatial,otieno2024public}.

Network spillovers, however, offer a broader perspective on interactions compared to spatial spillovers. Unlike spatial models, which rely on geographic proximity, network spillovers consider complex linkages between entities that are connected through various forms of relationships, such as similar economic indicators in the case of countries, industry sectors in the case of stocks, or any other source of shared characteristics, that is irrespective of the geographic location. Proximity-based networks extend, therefore, the concept of similarity by defining 'proximity' in a more abstract sense beyond geographic distance \citep[e.g.,][ derived proximity from financial balance sheet data]{fulle2024spatial}. In these networks, nodes that are proximate in terms of such similarities are likely to experience stronger volatility spillovers. By incorporating proximity-based networks into volatility models, we capture a richer and more nuanced view of how shocks propagate, offering critical insights into the broader dynamics of interconnected systems. Second, the network structure can be defined by the proximity/correlation of the observed time series at each node of the network \citep[e.g.,][]{diebold2014network,mattera2024network}. Thirdly, the network may be defined due to its physical nature, such as river networks \citep[e.g.,][]{tonkin2018role} or energy networks \citep[e.g.,][]{malinovskaya2023statistical}.

Despite spatial and/or network interactions, other approaches have been proposed to improve the fit of volatility models. For instance, empirical evidence in many fields suggests that better estimates of volatility dynamics are obtained by including additional latent factors. \cite{barigozzi2016generalized,barigozzi2017generalized,uddin2023risk} shows that in large panels of financial time series with dynamic factor structure on the levels or returns, the volatilities of the common and idiosyncratic components are exposed to the same volatility shocks. Therefore, estimating and including these latent factors in the model improves the estimation and the prediction of the volatility for the multivariate time series. In a different applicative domain, \cite{gorodnichenko2017level} identified separated factors for levels and volatility in macroeconomic time series, such as inflation and industrial production. Furthermore, \cite{christoffersen2019factor} provided evidence that the factor structure in daily commodity volatility is much stronger than the factor structure in returns for commodities.

In this study, we introduce a new dynamic spatiotemporal model to specify the volatility process of an outcome variable. We assume that the log volatility of the underlying outcome depends on the log-squared outcome of other entities in the network, the time lag of the log-squared outcome, and the time lag of the log-squared outcome of other entities in the network. We refer to the influence of other entities' log-squared outcomes as the spatial effect (the contemporaneous volatility spillover), the influence of the time-lag term as the temporal effect, and the influence of the time-lag of other entities' log-squared outcomes as the spatiotemporal effect (the lagged volatility spillover). In addition to these effects, we also account for the influence of observed and unobserved factors on the log-volatility process. To capture the impact of latent factors, we employ a factor structure that allows for entity-specific effects and permits arbitrary correlations between the latent and observed factors.

Our approach allows us to identify all these effects, and we propose an efficient estimation method based on the Bayesian MCMC framework. We begin by applying a log-squared transformation to derive an estimation equation that is linear in the parameters and the log-squared original error terms \citep{Taspinar:2021, Robinson:2009}. Following the time series literature \citep{Kim:1998, Omari:2007}, we approximate the distribution of the log-squared error terms in the transformed model using a finite mixture of normal distributions. To facilitate Bayesian estimation, we express the mixture density using an auxiliary latent random variable that serves as a mixture indicator. Finally, we augment the transformed model with the latent mixture indicators and suggest an efficient Gibbs sampler for estimation. Importantly, our approach enables us to recover the volatility estimates for each entity in the network over time as a natural by-product of the estimation process.

We recommend two Bayesian methods for determining the number of latent factors in the model. The first method is the deviance information criterion (DIC) proposed by \citet{David:2002}, which we demonstrate how to apply in our context. The DIC measures predictive performance, balancing model complexity and model fit. The second approach is based on the Bayesian Lasso, as considered by \citet{Park:2008}.  In this approach, we assume shrinkage priors for factor loadings to automatically exclude the latent factors that have less impact on the log-volatility process. In a simulation study, we show that our Bayesian estimation approach performs well and produces point estimates that are very close to true parameter values. 

In two empirical applications, we use our model to estimate the volatility processes of the US house price returns at the state level and the US stock market returns. Our results from both applications indicate the presence of economy-wide latent factors that affect the volatility of both the US house price returns and US stock market returns. For the US house returns, our findings show that the spatial, temporal, and spatiotemporal effects estimates are positive and statistically significant. For the US stock market returns, we find that the spatial and spatiotemporal effects estimates are positive and statistically significant, with relatively larger magnitudes. 

The rest of the paper is structured as follows. Section \ref{sec2} discusses the proposed dynamic spatiotemporal ARCH model with common factors. The interpretation of the model in terms of network process is therein discussed. Section \ref{sec3} shows the posterior analysis of the model, entering in the technicalities of the two algorithms proposed, while the simulation study is presented in detail in Section \ref{sec4}. Section \ref{sec5} provides two applications, one for house prices in the US and another one for the US stock market. While the first application adopts a purely spatial perspective, the second one shows the application of our method from a network perspective. Section~\ref{sec6} concludes with final remarks and future research directions.
 
\section{A dynamic spatiotemporal ARCH model with common factors}\label{sec2}

\noindent We consider the random process $\{Y_t(\mf{s}): \mf{s}\in \mf{D}_1\subseteq\mathbb{R}^d, d>1, t\in D_2\subseteq\mathbb{R}\}$, where $\mf{s}$ and $t$ denote the spatial location and time point, respectively. The data-type determines the structure of the spatial domain $\mf{D}_1$ and the time domain $D_2$, and we assume that $\mf{D}_1=\{\mf{s}_1,\hdots,\mf{s}_n\}$ and $D_2=\{1,2,\hdots,T\}$.  Then, we consider the following data generating process for $Y_t(\mf{s}_i)$:
\begin{align}
&Y_{t}(\mf{s}_i)=h_{t}^{1/2}(\mf{s}_i)\e_{t}(\mf{s}_i),\label{2.1}\\
&\log h_{t}(\mf{s}_i)=\sum_{j=1}^n\rho m_{ij}\log Y^2_{t}(\mf{s}_j)+\gamma\log Y^2_{t-1}(\mf{s}_j)+\sum_{j=1}^n\delta m_{ij}\log Y^2_{t-1}(\mf{s}_j)\label{2.2}\nonumber\\ 
&\quad\quad\quad\quad\quad+\mathbf{x}^{'}_{t}(\mf{s}_i)\bs{\beta}+\bs{\lambda}^{'}(\mf{s}_i)\mf{f}_t,
\end{align}
where $h_{t}(\mf{s}_i)$ is considered as the volatility term in location $\mf{s}_i$ at time $t$, and $\{\e_{it}(\mf{s}_i)\}$ is a sequence of i.i.d normal random variables with mean zero and unit variance. The log-volatility terms follow the process specified in \eqref{2.2},  where $\{m_{ij}\}$, for $i,j=1,\hdots,n$, are the non-stochastic weights specifying the strength of links among spatial units, $\mf{x}_{t}(\mf{s}_i)$ is a $k\times1$ vector of exogenous variables with the associated parameter vector $\bs{\beta}$, $\mf{f}_t$ is the $q\times1$ vector of unobserved common factors, and $\bs{\lambda}(\mf{s}_i)$ is the corresponding $q\times1$ vector of factor loadings. In the log-volatility equation, the spatial, temporal, and spatiotemporal effects of the log-squared outcome variable on the log-volatility are measured by the unknown scalar parameters $\gamma$, $\rho$, and $\delta$, respectively.  We assume that the initial value vector $\mf{Y}_0=(Y_{0}(\mf{s}_1),\hdots,Y_{0}(\mf{s}_n))^{'}$ is observable. 

The factor loadings and factors can be correlated with $\mf{x}_{t}(\mf{s}_i)$ in an arbitrary manner. This specification generalizes the dynamic spatiotemporal model suggested by \citet{Otto:2023} because the common factors specification $\bs{\lambda}^{'}(\mf{s}_i)\mf{f}_t$ can include the additive time and spatial fixed effects. For example, by setting $\mf{f}_t=(1,\alpha_{t0},0,\hdots,0)^{'}$ and $\bs{\lambda}(\mf{s}_i)=(\mu_0(\mf{s}_i),1,0\hdots,0)^{'}$, then we will obtain $\bs{\lambda}^{'}(\mf{s}_i)\mf{f}_t=\mu_0(\mf{s}_i)+\alpha_{t0}$, which is the additive structure assumed in \citet{Otto:2023}. Let $\bs{\Lambda}=(\bs{\lambda}_1,\hdots,\bs{\lambda}_n)^{'}$ be the $n\times q$ matrix of loadings. It is well known that the factors and loadings cannot be identified separately \citep{Bai:2003}. For example, for any orthogonal $q\times q$ matrix $\mf{Q}$, we have an observational equivalent relation $\bs{\Lambda}\mf{f}_t=\bs{\Lambda}\mf{Q}\mf{Q}^{'}\mf{f}_t$. Thus, to identify $\bs{\Lambda}$ and $\mf{f}_t$, we need to impose $q^2$ restrictions to pin down $\mf{Q}$. Our ensuing analysis will show that the estimation of the log-volatility term and other parameters only requires the estimation of the product $\bs{\Lambda}\mf{f}_t$. That is, our analysis does not require the identification of $\bs{\Lambda}$ and $\mf{f}_t$ separately.

\begin{figure}[!htb]
	\begin{center}
	    \includegraphics[width=0.8\textwidth]{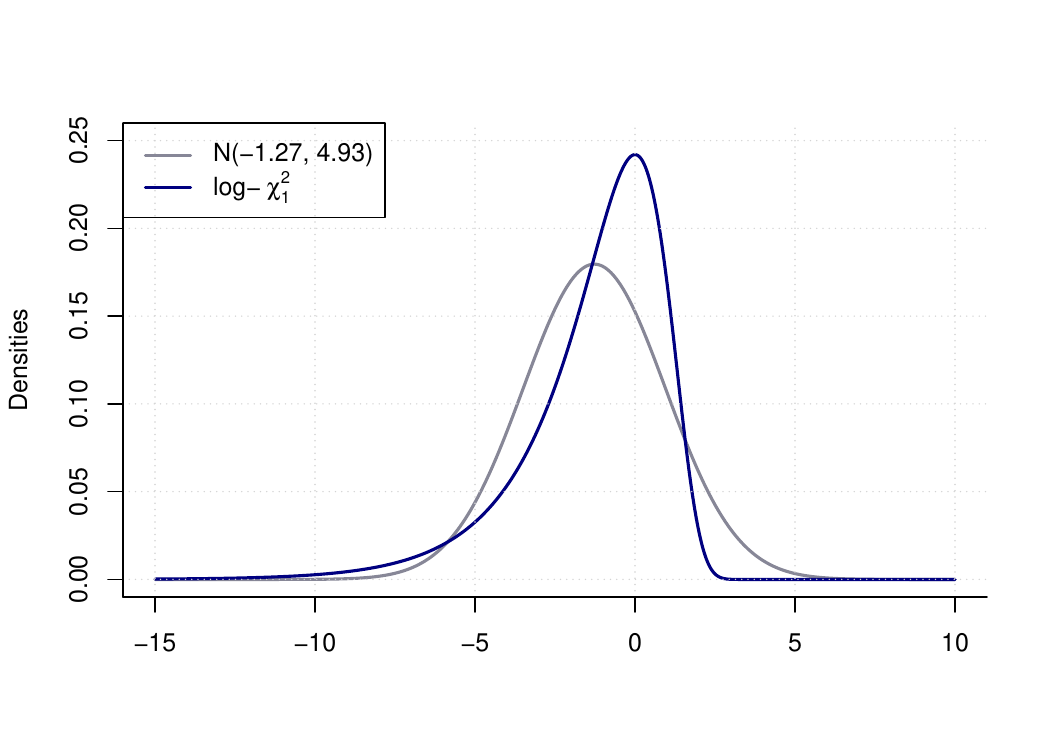}
	\end{center}
	\vspace{-0.80cm}
	\caption{The density plots of $N(-\vartheta - \log(2),\,\pi^2/2)$ and  $\log\chi^2_1$}
	\label{fig:logchi}
\end{figure}

Define $Y^*_{t}(\mf{s}_i)=\log Y^2_{t}(\mf{s}_i)$, $h^{*}_{t}(\mf{s}_i)=\log h_{t}(\mf{s}_i)$ and $\e^*_{t}(\mf{s}_i)=\log\e^2_{t}(\mf{s}_i)$. Then, we apply the log-squared transformation to \eqref{2.1} and obtain
\begin{align}\label{2.3}
Y^{*}_{t}(\mf{s}_i)=h^{*}_{t}(\mf{s}_i)+\e^{*}_{t}(\mf{s}_i),
\end{align}
Using vector notation, we can express \eqref{2.2} and \eqref{2.3}  as
\begin{align}
&\mf{Y}^{*}_t=\mf{h}^{*}_t+\bs{\e}^{*}_t,\label{2.4}\\
&\mf{h}^{*}_t=\rho\mf{M}\mf{Y}^{*}_t+\gamma \mf{Y}^{*}_{t-1}+\delta\mf{M}\mf{Y}^{*}_{t-1}+\mf{X}_t\bs{\beta}+\bs{\Lambda}\mf{f}_t,\label{2.5}
\end{align}
where $\mf{M}=(m_{ij})$ is the $n\times n$ network matrix, $\mf{Y}^{*}_t=(Y^{*}_{t}(\mf{s}_1),\hdots,Y^{*}_{t}(\mf{s}_n))^{'}$, $\mf{h}^{*}_t=(h^{*}_{t}(\mf{s}_1),\hdots,h^{*}_{t}(\mf{s}_n))^{'}$, $\bs{\e}^{*}_t=(\e^{*}_{t}(\mf{s}_1),\hdots,\e^{*}_{t}(\mf{s}_n))^{'}$, and $\mf{X}_t=(\mf{x}_{t}(\mf{s}_1),\hdots,\mf{x}_{t}(\mf{s}_n))^{'}$.  Then, we obtain an estimation equation by substituting \eqref{2.5} into \eqref{2.4}: 
\begin{align}\label{2.6}
\mf{Y}^{*}_t=\rho\mf{M}\mf{Y}^{*}_t+\gamma \mf{Y}^{*}_{t-1}+\delta\mf{M}\mf{Y}^{*}_{t-1}+\mf{X}_t\bs{\beta}+\bs{\Lambda}\mf{f}_t+\bs{\e}^{*}_t.
\end{align}
The estimation equation in \eqref{2.2} is in the form of a spatial dynamic panel data model with common factors, as considered by \citet{Shi:2017} and \citet{Bai:2021}, with two notable new features. Firstly, the outcome variable, the spatial lag term ($\mf{M}\mf{Y}^{*}_t$),  and the spatiotemporal lag term ($\mf{M}\mf{Y}^{*}_{t-1}$)  are formulated in terms of the log-squared outcome variable. Secondly, the elements of $\bs{\e}^{*}_t$ are the log-squared original disturbance terms, i.e., $\e^*_{t}(\mf{s}_i)=\log\e^2_{t}(\mf{s}_i)$. If we assume that $\e_{t}(\mf{s}_i)\sim N(0,1)$, then $\e^{*}_{t}(\mf{s}_i)\sim\log\chi^2_1$, which is the log-chi squared distribution with one degree of freedom with the following density:\footnote{In our ensuing analysis, we use $p(\cdot)$ to denote density functions and omit $\mf{X}_t$ in the conditional set for the sake of simplicity.}
\begin{align}\label{2.7} 
p(\e^{*}_{t}(\mf{s}_i))=\frac{1}{\sqrt{2\pi}}\exp\left(-\frac{1}{2}\left(\exp(\e^{*}_{t}(\mf{s}_i))-\e^{*}_{t}(\mf{s}_i)\right)\right),\quad-\infty<\e^{*}_{t}(\mf{s}_i)<\infty,
\end{align}
The first two moments of $\log\chi^2_1$ are $\E(\e^{*}_{t}(\mf{s}_i)) = -\vartheta - \log(2) \approx-1.2704$, where $\vartheta$ is Euler's constant, and $\text{Var}(\e^{*}_{t}(\mf{s}_i))=\pi^2/2\approx4.9348$ \citep[pp. 379-380]{Peter:2012}. In Figure~\ref{fig:logchi}, we compare this density with that of $N(-\vartheta - \log(2),\,\pi^2/2)$. As seen from the figure, the log-chi squared distribution exhibits significant skewness with a long left tail. 

Let $\mf{S}(\rho)=\mf{I}_n-\rho\mf{M}$ and $\bs{\varphi}=(\gamma,\delta)^{'}$. Then, under the assumption that $\mf{S}^{-1}(\rho)$ exists,  the reduced form of the model can be expressed as 
\begin{align}\label{2.8}
\mf{Y}^{*}_t=\mf{A}(\rho,\bs{\varphi})\mf{Y}^{*}_{t-1}+\mf{S}^{-1}(\rho)\mf{X}_t\bs{\beta}+\mf{S}^{-1}(\rho)\bs{\Lambda}\mf{f}_t+\mf{S}^{-1}(\rho)\bs{\e}^{*}_t,
\end{align}
where $\mf{A}(\rho,\bs{\varphi})=\mf{S}^{-1}(\rho)\left(\gamma\mf{I}_n+\delta\mf{M}\right)$. Thus, if all eigenvalues of $\mf{A}(\rho,\bs{\varphi})$ lie within the unit circle, then $\mf{Y}^{*}_t$ will have a stable process \citep[p. 259]{Hamilton:1994}. By the spectral radius theorem, $\mf{Y}^{*}_t$ has a stable process if $\Vert\mf{S}(\rho)\Vert<1$ and $\Vert\mf{A}(\rho,\bs{\varphi})\Vert<1$, where $\Vert\cdot\Vert$ is any matrix norm. If we use the matrix row sum norm and assume that $\mf{M}$ is row normalized, then it can be shown that a sufficient condition for stability is $|\rho|+|\gamma|+|\delta|<1$ \citep{Yu:2008}. 

We note that our specification can be applied to model more general data observed across (random) networks rather than exclusively spatio-temporal data \citep[see, e.g.,][]{fulle2024spatial, mattera2024network}.  A network $G$ is defined as the tuple $G=(V,E)$, where $V$ is a set of $n$ distinct vertices/nodes that are connected by directed or undirected edges, comprised in the set $E$. In practice, the edges are often represented by a (potentially sparse) $n$-dimensional adjacency matrix. In addition, the edges could be weighted to accommodate different ``lengths/distances'' between the nodes. Let $V=\mathbf{D}_1$ and define $\{m_{ij}\}$ the network weights representing the strength of the connection between two nodes $i$ and $j$. These weights can either be known in advance or estimated based on cross-sectional features \citep{fulle2024spatial} or temporal similarities \citep{mattera2024network}. In this case, we refer to the process as Network log-ARCH with common factors. While the technical details remain the same in the case of a network structure, the choice of different network definitions influences the interpretation of the results. This idea could also be extended to random graphs with a fixed set of nodes (potentially including isolated nodes), where the weight matrices $\mf{M}$ are allowed to vary over time.

\section{Posterior Analysis}\label{sec3}

\noindent For posterior analysis, we assume the following independent priors:
\begin{align}\label{3.8}
&\rho\sim\text{Uniform}(-1,1),\,\bs{\varphi}|\mf{b}_{\varphi},\mf{B}_{\varphi}\sim N(\mf{b}_{\varphi},\mf{B}_{\varphi}),\, \bs{\beta}|\mf{b}_{\beta},\mf{B}_{\beta}\sim N(\mf{b}_{\beta},\mf{B}_{\beta}), \nonumber\\ 
&\bs{\lambda}(\mf{s}_i)|\mf{b}_{\lambda},\mf{B}_{\lambda}\sim N(\mf{b}_{\lambda},\mf{B}_{\lambda}), \,i=1,\hdots,n,\, \mf{f}_t\sim N(\mf{0},\mf{I}_q), \,t=1,\hdots,T,
\end{align}
where $\text{Uniform}(c_1,\,c_2)$ is the uniform distribution over the interval $(c_1,\,c_2)$. When $\mf{M}$ is row normalized, $\mf{S}^{-1}(\rho)$ exists for $\rho\in(-1,1)$. Therefore, we assume that $\rho\sim\text{Uniform}(-1,1)$. When $\mf{M}$ is not row-normalized, $\mf{S}^{-1}(\rho)$ exists for $\rho\in(-1/\tau,1/\tau)$, where $\tau$ is the spectral radius of $\mf{M}$ \citep{KP:2010}. Thus, we can alternatively assume that $\rho\sim\text{Uniform}(-1/\tau,1/\tau)$. Additionally, we also require that the prior distributions of $\rho$ and $\bs{\varphi}$ are subject to the stability condition stated in Section~\ref{sec2}. The prior distributions assumed for $\varphi$ and $\bs{\beta}$ are commonly used priors in the literature. Finally, following \citet{Geweke:1996} and \citet{Han:2016}, we assume that $\mf{f}_t\sim N(\mf{0},\mf{I}_q)$, where $\mf{I}_q$ is the $q\times q$ identity matrix.

Following the time series literature on the standard stochastic volatility models \citep{Kim:1998, Omari:2007}, we use a finite mixture of normal distributions to approximate $p(\e^*_{t}(\mf{s}_i))$:
\begin{align}\label{3.3}
p(\e^*_{t}(\mf{s}_i))\approx\sum_{j=1}^{10}p_j\times \phi(\e^*_{t}(\mf{s}_i)|\mu_j,\,\sigma^2_j),
\end{align}
where $\phi(\e^*_{t}(\mf{s}_i)|\mu_j,\,\sigma^2_j)$ denotes the Gaussian density function with mean $\mu_j$ and variance $\sigma^2_j$,  and $p_j$ is the probability of $j$th mixture component.  The parameters of this mixture distribution are given in Table~\ref{table1}. These parameters are chosen by matching the first four moments of the ten-component normal mixture distribution with that of $p(\e^*_{t}(\mf{s}_i))$. Under \eqref{3.3}, our model becomes a mixture of normal distributions. Furthermore, this approach does not pose any estimation difficulties, as the parameters of the mixture distribution are pre-determined. 

\definecolor{LightCyan}{rgb}{0.88,1,1}
\begin{table}[!htb]
\begin{center}
\caption{Parameters of ten-component normal mixture distribution} 
\label{table1} 
\setlength{\tabcolsep}{5pt} 
\renewcommand{\arraystretch}{1} 
\begin{tabular*}{0.65\textwidth}{@{\extracolsep{\fill} }ccrc} 
\hline\hline
Components& $p_j$ &\multicolumn{1}{c}{$\mu_j$}&$\sigma^2_j$ \\
\hline
1 &0.00609 &1.92677& 0.11265 \\ 
2&0.04775& 1.34744& 0.17788\\
3 &0.13057& 0.73504& 0.26768\\
4&0.20674 &0.02266& 0.40611\\
5 &0.22715&-0.85173 &0.62699\\
6&0.18842&-1.97278& 0.98583\\
7&0.12047&-3.46788& 1.57469\\
8 &0.05591&-5.55246& 2.54498\\
9&0.01575 &-8.68384 &4.16591\\
10&0.00115&-14.65000&7.33342\\
\hline\hline
\end{tabular*}
\end{center}
\end{table}

\noindent Let $Z_{t}(\mf{s}_i)$ be the mixture component indicator variable that takes values from $\{1,2,\hdots,10\}$. Then, using $Z_{t}(\mf{s}_i)$, we can express the finite mixture distribution in \eqref{3.3} as
\begin{align}\label{4.6}
\e^*_{t}(\mf{s}_i)|(Z_{t}(\mf{s}_i)=j)\sim N(\mu_j,\,\sigma^2_j),\quad\text{and}\quad \mathbb{P}(Z_{t}(\mf{s}_i)=j)=p_j,\quad j=1,2,\hdots,10,
\end{align}
where $\mathbb{P}(Z_{t}(\mf{s}_i)=j)=p_j$ is the probability that $Z_{t}(\mf{s}_i)$ equals $j$.  Let $\mathbf{Z}_t=(Z_{t}(\mf{s}_1),\hdots,Z_{t}(\mf{s}_n))^{'}$, $\mathbf{d}_t=(\mu_{Z_{t}(\mf{s}_1)},\hdots,\mu_{Z_{t}(\mf{s}_n)})^{'}$ and $\bs{\Sigma}_t=\Diag(\sigma^2_{Z_{t}(\mf{s}_1)},\hdots,\sigma^2_{Z_{t}(\mf{s}_n)})$, where $\Diag(a_1,\hdots,a_n)$ denotes the $n\times n$ diagonal matrix with the $i$th diagonal element $a_i$. Then, the mixture component indicator variable allows us to determine $\bs{\e}^*_t|\bs{Z}_t\sim N(\mf{d}_t,\,\bs{\Sigma}_t)$. Thus, we can express the likelihood function of \eqref{2.6} as
\begin{align}\label{3.5}
p(\mf{Y}^{*}_t|\mf{Z}_t,\bs{\beta},\bs{\Lambda},\bs{f}_t,\bs{\varphi},\rho)&=(2\pi)^{-n/2}\left|\mf{S}(\rho)\right|\left(\prod_{i=1}^n\sigma^2_{Z_t(\mf{s}_i)}\right)^{-1/2}\nonumber\\
&\times\exp\left(-\frac{1}{2}\left(\mf{S}(\rho)\mf{Y}^{*}_t-\mf{b}_{Y_t}\right)^{'}\bs{\Sigma}^{-1}_t\left(\mf{S}(\rho)\mf{Y}^{*}_t-\mf{b}_{Y_t}\right)\right),
\end{align}
where $\mf{b}_{Y_t}=\mf{W}_t\bs{\varphi}+\mf{X}_t\bs{\beta}+\bs{\Lambda}\mf{f}_t+\mf{d}_t$ with $\mf{W}_t=(\mf{Y}^{*}_{t-1},\mf{M}\mf{Y}^{*}_{t-1})$. Let $\mf{Y}^{*}=(\mf{Y}^{*'}_1,\hdots,\mf{Y}^{*'}_T)^{'}$, $\mf{Z}=(\mf{Z}^{'}_1,\hdots,\mf{Z}^{'}_T)^{'}$, and $\mf{F}=(\mf{f}^{'}_1,\hdots,\mf{f}^{'}_T)^{'}$. Then, the joint posterior distribution $p(\mf{Z},\bs{\beta},\bs{\Lambda},\mf{F},\bs{\varphi},\rho|\mf{Y}^{*})$ can be expressed as
\begin{align}
p(\mf{Z},\bs{\beta},\bs{\Lambda},\mf{F},\bs{\varphi},\rho|\mf{Y}^{*})\propto p(\mf{Y}^{*}|\mf{Z},\bs{\beta},\bs{\Lambda},\mf{F},\bs{\varphi}) p(\mf{Z}) p(\bs{\beta}) p(\bs{\Lambda}) p(\mf{F}) p(\bs{\varphi})p(\rho),
\end{align}
where $p(\mf{Y}^{*}|\mf{Z},\bs{\beta},\bs{\Lambda},\mf{F},\bs{\varphi},\rho)=\prod_{t=1}^Tp(\mf{Y}^{*}_t|\mf{Z}_t,\bs{\beta},\bs{\Lambda},\bs{f}_t,\bs{\varphi},\rho)$ and $p(\mf{a})$ denotes the prior distribution of $\mf{a}$  for $\mf{a}\in \{\mf{Z},\bs{\beta},\bs{\Lambda},\mf{F},\bs{\varphi},\rho\}$. We note that only the conditional posterior distribution of $\rho$ does not take a known form because the likelihood function is non-linear in $\rho$ as shown in \eqref{3.5}. For this parameter, we resort to the random walk Metropolis-Hastings (MH) algorithm suggested by \citet{Lesage:2009}.\footnote{It is also possible to sample $\rho$ and $\bs{\varphi}$ jointly by using the adaptive MH algorithm considered in \citet{Haario:2001} and \citet{Roberts:2009}. The main feature of the adaptive MH algorithm is that historical draws are used to determine the mean and covariance of the proposal distribution. \citet{Han:2016} and \citet{Yang:2023} use this algorithm for static and dynamic spatial panel data models.} The conditional posterior distributions for the remaining parameters take known forms and can be determined by the standard Bayesian analysis used for a linear regression model. We summarize how these distributions can be determined in  Algorithm~\ref{a1}.
\begin{algorithm}[Standard MCMC]\label{a1}
\leavevmode   \normalfont
\begin{enumerate}
\item Sampling step for $\mf{Z}$: Let $\mf{Y}_{zt}=\mf{S}(\rho)\mf{Y}^{*}_t-\mf{W}_t\bs{\varphi}-\mf{X}_t\bs{\beta}-\bs{\Lambda}\mf{f}_t$ with the $(\mf{s}_i,t)$th element denoted by $Y_{zt}(\mf{s}_i)$. Then, we have $\mf{Y}_{zt}|\mf{Z}_t\sim N(\mf{d}_t,\bs{\Sigma}_t)$. Thus, the discrete conditional posterior distribution of $\mf{Z}_t$ can be determined from $p(\mf{Z}_t|\mf{Y}_{zt})\propto p(\mf{Y}_{zt}|\mf{Z}_t)p(\mf{Z}_t)=\prod_{i=1}^{n}p(Y_{zt}(\mf{s}_i)|Z_t(\mf{s}_i))p(Z_{t}(\mf{s}_i))$, which yields
\begin{align}
p(Z_t(\mf{s}_i)=j|Y_{zt}(\mf{s}_i))=
\frac{p_j\phi(Y_{zt}(\mf{s}_i)|\mu_j,\,\sigma^2_j)}{\sum_{k=1}^{10}p_k\phi(Y_{zt}(\mf{s}_i)|\mu_k,\,\sigma^2_k)},\,j=1,\hdots,10,\,i=1,\hdots,n,
\end{align} 
for $t=1,\hdots,T$, where the denominator is the normalization constant.
\item Sampling step for $\bs{\beta}$: Let $\mf{Y}_{\beta t}=\mf{S}(\rho)\mf{Y}^{*}_t-\mf{W}_t\bs{\varphi}-\bs{\Lambda}\mf{f}_t-\mf{d}_t$. Then, from $p(\bs{\beta}|\mf{Y}^{*},\mf{Z},\bs{\Lambda},\mf{F},\bs{\varphi},\rho)\propto p(\mf{Y}^{*}|\mf{Z},\bs{\beta},\bs{\Lambda},\mf{F},\bs{\varphi},\rho) p(\bs{\beta})$, we obtain
\begin{align}
\bs{\beta}|\mf{Y}^{*},\mf{Z},\bs{\Lambda},\mf{F},\bs{\varphi},\rho\sim N(\bs{\mu}_\beta,\mf{K}_{\beta}),
\end{align}
where $\mf{K}_{\beta}=\left(\mf{B}^{-1}_{\beta}+\sum_{t=1}^T\mf{X}^{'}_t\bs{\Sigma}^{-1}_t\mf{X}_t\right)^{-1}$ and $\bs{\mu}_{\beta}=\mf{K}_{\beta}\left(\mf{B}^{-1}_{\beta}\mf{b}_{\beta}+\sum_{t=1}^T\mf{X}^{'}_t\bs{\Sigma}^{-1}_t\mf{Y}_{\beta t}\right)$.
\item Sampling step for $\mf{F}$: Let $\mf{Y}_{ft}=\mf{S}(\rho)\mf{Y}^{*}_t-\mf{W}_t\bs{\varphi}-\mf{X}_t\bs{\beta}-\mf{d}_t$. Then, $p(\mf{F}|\mf{Y}^{*},\mf{Z},\bs{\beta},\bs{\Lambda},\bs{\varphi},\rho)\propto p(\mf{Y}^{*}|\mf{Z},\bs{\beta},\bs{\Lambda},\mf{F},\bs{\varphi},\rho) p(\mf{F})$ yields
\begin{align}
\mf{f}_t|\mf{Y}^{*},\mf{Z},\bs{\beta},\bs{\Lambda},\bs{\varphi},\rho\sim N\left(\bs{\mu}_{f_t},\mf{K}_{f_t}\right),\quad t=1,\hdots,T,
\end{align}
where $\mf{K}_{f_t}=\left(\mf{I}_q+\bs{\Lambda}^{'}\bs{\Sigma}^{-1}_t\bs{\Lambda}\right)^{-1}$ and $\bs{\mu}_{f_t}=\mf{K}_{f_t}\bs{\Lambda}^{'}\bs{\Sigma}^{-1}_t\mf{Y}_{ft}$.
\item Sampling step for $\bs{\Lambda}$: Let $Y_{ft}(\mf{s}_i)$ be the $(\mf{s}_i,t)$th element of $\mf{Y}_{ft}$. Then, from $p(\mf{\Lambda}|\mf{Y}^{*},\mf{Z},\bs{\beta},\mf{F},\bs{\varphi},\rho)\propto p(\mf{Y}^{*}|\mf{Z},\bs{\beta},\bs{\Lambda},\mf{F},\bs{\varphi},\rho) p(\bs{\Lambda})$, we obtain
\begin{align}
\bs{\lambda}(\mf{s}_i)|\mf{Y}^{*},\mf{Z},\bs{\beta},\mf{F},\bs{\varphi},\rho\sim N\left(\bs{\mu}_{\lambda_i},\mf{K}_{\lambda_i}\right),\quad i=1,\hdots,n,
\end{align}
where $\mf{K}_{\lambda_i}=\left(\mf{B}^{-1}_{\lambda}+\sum_{t=1}^T\frac{1}{\sigma^2_{Z_t(\mf{s}_i)}}\mf{f}_t\mf{f}^{'}_t\right)^{-1}$ and  $\bs{\mu}_{\lambda_i}=\mf{K}_{\lambda_i}\left(\mf{B}^{-1}_{\lambda}\mf{b}_{\lambda}+\sum_{t=1}^T\frac{1}{\sigma^2_{Z_t(\mf{s}_i)}}\mf{f}_tY_{ft}(\mf{s}_i)\right)$.
\item Sampling step for $\bs{\varphi}$: Let $\mf{Y}_{\varphi t}=\mf{S}(\rho)\mf{Y}^{*}_t-\mf{X}_t\bs{\beta}-\bs{\Lambda}\mf{f}_t-\mf{d}_t$. Then, from $p(\bs{\varphi}|\mf{Y}^{*},\mf{Z},\bs{\beta},\bs{\Lambda},\mf{F},\rho)\propto p(\mf{Y}^{*}|\mf{Z},\bs{\beta},\bs{\Lambda},\mf{F},\bs{\varphi},\rho) p(\bs{\varphi})$, we obtain
\begin{align}
\bs{\varphi}|\mf{Y}^{*},\mf{Z},\bs{\beta},\bs{\Lambda},\mf{F},\rho\sim N(\bs{\mu}_{\varphi},\mf{K}_{\varphi}),
\end{align}
where $\mf{K}_{\varphi}=\left(\mf{B}^{-1}_{\varphi}+\sum_{t=1}^T\mf{W}^{'}_t\bs{\Sigma}^{-1}_t\mf{W}_t\right)^{-1}$ and $\bs{\mu}_{\varphi}=\mf{K}_{\varphi}\left(\mf{B}^{-1}_{\varphi}\mf{b}_{\varphi}+\sum_{t=1}^T\mf{W}^{'}_t\bs{\Sigma}^{-1}_t\mf{Y}_{\varphi t}\right)$.
\item Sampling step for $\rho$: Let $\mf{a}^{(g)}$ be the draw generated at the $g$th sampling step for $\mf{a}\in \{\mf{Z},\bs{\beta},\bs{\Lambda},\mf{F},\bs{\varphi},\rho\}$.  At the $g$th sampling step, generate a candidate value $\tilde{\rho}$ according to $\tilde{\rho}=\rho^{(g-1)}+c_{\rho}\times N(0,1)$, where $c_{\rho}$ is a tuning parameter.\footnote{The tuning parameter can be determined during the estimation such that the acceptance rate falls between $40\%$ and $60\%$ \citep{Lesage:2009}.} Then, compute the following acceptance probability:
\begin{align*}
\mathbb{P}(\rho^{(g-1)}, \tilde{\rho})=\min\left(\frac{p(\mf{Y}^{*}|\mf{Z}^{(g)},\bs{\beta}^{(g)},\bs{\Lambda}^{(g)},\mf{F}^{(g)},\bs{\varphi}^{(g)},\tilde{\rho})}{p(\mf{Y}^{*}|\mf{Z}^{(g)},\bs{\beta}^{(g)},\bs{\Lambda}^{(g)},\mf{F}^{(g)},\bs{\varphi}^{(g)},\rho^{(g-1)})},\,1\right).
\end{align*}
Return $\tilde{\rho}$  with probability $\mathbb{P}(\rho^{(g-1)}, \tilde{\rho})$, otherwise return $\rho^{(g-1)}$ .
 
\end{enumerate}
\end{algorithm}
In Algorithm~\ref{a1}, the factors and loadings enter  the conditional posterior distributions of $\bs{\beta}$, $\bs{\varphi}$ and $\rho$ as a product term $\mf{\Lambda}\mf{f}_t$. Moreover, the volatility estimates can be recovered from $\hat{\mf{h}}^{*}_t=\hat{\rho}\mf{M}\mf{Y}^{*}_t+\hat{\gamma} \mf{Y}^{*}_{t-1}+\hat{\delta}\mf{M}\mf{Y}^{*}_{t-1}+\mf{X}_t\hat{\bs{\beta}}+\hat{\bs{\Lambda}\mf{f}}_t$, where $\hat{\rho}$, $\hat{\gamma}$, $\hat{\delta}$, $\hat{\beta}$ and  $\hat{\bs{\Lambda}\mf{f}}_t$ are the estimated posterior means obtained through Algorithm~\ref{a1}. Thus, our estimation approach does not require the identification of $\mf{f}_t$ and $\bs{\Lambda}$ separately.

The Gibbs sampler in Algorithm~\ref{a1} requires that the number of factors $q$ is known. The literature suggests several ways that can be used to determine the number of factors \citep{Ng:2002, Alexei:2010}. Following \citet{Han:2016}, we suggest using the deviance information criterion (DIC) suggested by \citet{David:2002} to determine the number of factors. The DIC can be considered as the Bayesian version of AIC and take the following from:
\begin{align}\label{3.14}
\text{DIC}=-4\E\left(\ln p(\mf{Y}^{*}|\mf{Z},\bs{\beta},\bs{\Lambda},\mf{F},\bs{\varphi},\rho)|\mf{Y}^{*}\right)+2\ln p(\mf{Y}^{*}|\bar{\mf{Z}},\bar{\bs{\beta}},\bar{\bs{\Lambda}},\bar{\mf{F}},\bar{\bs{\varphi}},\bar{\rho}),
\end{align}
where the expectation is taken with respect to the joint posterior distribution and $\bar{\mf{a}}\in\{\bar{\mf{Z}},\bar{\bs{\beta}},\bar{\bs{\Lambda}},\bar{\mf{F}},\bar{\bs{\varphi}},\bar{\rho}\}$ denotes the posterior mean. Let $\{\mf{Z}^{(g)},\bs{\beta}^{(g)},\bs{\Lambda}^{(g)},\mf{F}^{(g)},\bs{\varphi}^{(g)},\rho^{(g)}\}_{g=1}^R$ be a sequence of posterior draws generated through Algorithm~\ref{a1}. Then, we can compute  $\E\left(\ln p(\mf{Y}^{*}|\mf{Z},\bs{\beta},\bs{\Lambda},\mf{F},\bs{\varphi},\rho)|\mf{Y}^{*}\right)$ in the DIC measure by $\E\left(\ln p(\mf{Y}^{*}|\mf{Z},\bs{\beta},\bs{\Lambda},\mf{F},\bs{\varphi},\rho)|\mf{Y}^{*}\right)\approx\frac{1}{R}\sum_{g=1}^R\ln p(\mf{Y}^{*}|\mf{Z}^{(g)},\bs{\beta}^{(g)},\bs{\Lambda}^{(g)},\mf{F}^{(g)},\bs{\varphi}^{(g)},\rho^{(g)})$. The second term $\ln p(\mf{Y}^{*}|\bar{\mf{Z}},\bar{\bs{\beta}},\bar{\bs{\Lambda}},\bar{\mf{F}},\bar{\bs{\varphi}},\bar{\rho})$ in \eqref{3.14} is simply the log-likelihood function evaluated at the posterior means. The DIC is a measure of predictive performance. Thus, a model with the smallest DIC value performs relatively better in terms of predictive accuracy. It is important to note that the DIC also does not require the separate identification of $\mf{f}_t$ and $\bs{\Lambda}$, as the factors and factor loadings are incorporated as a product term in the likelihood function.

An alternative approach for determining the number of factors can be based on a Bayesian Lasso approach, which can be very useful when there are many factors. In the context of a linear regression model, \citet{Tibshirani:1996} show that the Lasso coefficient estimates can be interpreted as the Bayesian posterior mode estimates under the independent Laplace prior distributions for the coefficients. Using this connection, \citet{Park:2008} develop a Gibbs sampler by assuming independent Laplace prior distributions for the coefficients. The sampler is based on the property that a Laplace distribution can be represented as a scale mixture of normal distribution, where the scale variable has an exponential distribution \citep{Andrews:1974}. Let $\lambda_m(\mf{s}_i)$ be the $m$th element of $\bs{\lambda}(\mf{s}_i)$ for $m=1,\hdots,q$.  Following  \citet{Park:2008}, we assume the following Laplace distribution for $\bs{\lambda}(\mf{s}_i)$:
\begin{align}\label{3.12}
p(\bs{\lambda}(\mf{s}_i))\propto\prod_{m=1}^q\frac{\phi}{2}\exp(-\phi|\lambda_m(\mf{s}_i)|),
\end{align}
where $\phi\geq0$ is the shrinkage parameter. In order to express this prior distribution in terms of a scale normal distribution, we use the following identity \citep{Andrews:1974}:
\begin{align}\label{3.13}
\frac{c}{2}\exp(-c|h|)=\int_{0}^{\infty}\frac{1}{\sqrt{2\pi a}}\exp(-\frac{h^2}{2a})\frac{c^2}{2}\exp(-\frac{c^2a}{2})\text{d}a.
\end{align}
In \eqref{3.13}, if we set $c=\phi$, $h=\lambda_m(\mf{s}_i)$, and $a=\tau^2_m$, where $\tau^2_m$ is an hyper-parameter for $\lambda_m(\mf{s}_i)$, we obtain
\begin{align}\label{3.14}
\frac{\phi}{2}\exp(-\phi|\lambda_m(\mf{s}_i)|)=\int_{0}^{\infty}\frac{1}{\sqrt{2\pi\tau^2_m}}\exp(-\frac{\lambda^2_m(\mf{s}_i)}{2\tau^2_m})\frac{\phi^2}{2}\exp(-\frac{\phi^2\tau^2_m}{2})\text{d}\tau^2_m.
\end{align}
The right hand side of \eqref{3.14} suggests that $\lambda_m(\mf{s}_i)|\tau^2_m\sim N(0,\tau^2_m)$ and $\tau^2_m|\phi^2\sim\text{Exp}(\phi^2/2)$, where $\text{Exp}(a)$ denotes the exponential distribution with rate $a$. Let $\bs{\tau}^2=(\tau^2_1,\hdots,\tau^2_q)^{'}$ and $\mf{D}_{\tau}=\text{Diag}(\tau^2_1,\hdots,\tau^2_q)$. Then, following  \citet{Park:2008} and \citet{Casella:2010}, we assume the following prior distributions:
\begin{align}
&\bs{\lambda}(\mf{s}_i)|\bs{\tau}^2\sim N(\mf{0},\mf{D}_{\tau}),\, i=1,\hdots,n,\quad \phi^2|c,d\sim \text{Gamma}(c,d),\nonumber\\
&\tau^2_m|\phi^2\sim \text{Exp}(\phi^2/2),\,m=1,\hdots,q,
\end{align}
where $\text{Gamma}(c,d)$ denotes the gamma distribution with shape parameter $c$ and the scale parameter $d$. This approach will require two new sampling steps for $\bs{\tau}^2$ and $\phi^2$, and the adjustment of the sampling step for $\bs{\lambda}(\mf{s}_i)$. We summarize this shrinkage MCMC approach in Algorithm~\ref{a2}.
\begin{algorithm}[Bayesian Lasso, shrinkage MCMC]\label{a2}
\leavevmode   \normalfont
\begin{enumerate}
\item The sampling steps for $\mf{Z}$, $\bs{\beta}$, $\mf{F}$, $\bs{\varphi}$, and $\rho$ remain the same as those in Algorithm~\ref{a1}.
\item Sampling step for $\bs{\Lambda}$: Let $Y_{ft}(\mf{s}_i)$ be the $(\mf{s}_i,t)$th element of $\mf{Y}_{ft}$. Then, from $p(\mf{\Lambda}|\mf{Y}^{*},\mf{Z},\bs{\beta},\mf{F},\bs{\varphi},\rho,\bs{\tau}^2)\propto p(\mf{Y}^{*}|\mf{Z},\bs{\beta},\bs{\Lambda},\mf{F},\bs{\varphi},\rho) p(\bs{\Lambda}|\bs{\tau}^2)$, we obtain
\begin{align}
\bs{\lambda}(\mf{s}_i)|\mf{Y}^{*},\mf{Z},\bs{\beta},\mf{F},\bs{\varphi},\rho\sim N\left(\bs{\mu}_{\lambda_i},\mf{K}_{\lambda_i}\right),\quad i=1,\hdots,n,
\end{align}
where $\mf{K}_{\lambda_i}=\left(\mf{D}^{-1}_{\tau}+\sum_{t=1}^T\frac{1}{\sigma^2_{Z_t(\mf{s}_i)}}\mf{f}_t\mf{f}^{'}_t\right)^{-1}$ and  $\bs{\mu}_{\lambda_i}=\mf{K}_{\lambda_i}\left(\sum_{t=1}^T\frac{1}{\sigma^2_{Z_t(\mf{s}_i)}}\mf{f}_tY_{ft}(\mf{s}_i)\right)$.
 \item Sampling step for $\bs{\tau}^2$: First, note that $\frac{1}{\tau^2_m}|\phi^2\sim \frac{\phi^2}{2}\tau^4_m\exp(-\frac{\phi^2\tau^2_m}{2})$. Then, from $p(\frac{1}{\tau^2_m}|\bs{\Lambda},\phi^2)\propto p(\bs{\Lambda}|\bs{\tau}^2) p(\frac{1}{\tau^2_m}|\phi^2)$, we obtain
 \begin{align}
 \frac{1}{\tau^2_m}|\bs{\Lambda},\phi^2\sim\text{Inverse-Gaussian}\left(\sqrt{\frac{\sum_{i=1}^n\lambda^2_m(\mf{s}_i)}{\phi^2}}, \sum_{i=1}^n\lambda^2_m(\mf{s}_i)\right),\,m=1,\hdots,q,
 \end{align}
 where $\text{Inverse-Gaussian}\left(a,b\right)$ is the inverse Gaussian distribution with mean $a$ and shape parameter $b$.
\item Sampling step for $\phi^2$: From $p(\phi^2|\bs{\tau}^2)\propto p(\bs{\tau}^2|\phi^2)p(\phi^2)$, we obtain
\begin{align}
\phi^2|\bs{\tau}^2\sim\text{Gamma}\left(c+q,\,d+\frac{1}{2}\sum_{m=1}^q\tau^2_m\right).
\end{align}

\end{enumerate}

\end{algorithm}

\section{Simulation Evidence}\label{sec4}

In this section, we consider a simulation exercise to elucidate the finite sample performance of the suggested Gibbs samplers with and without regularization. To this end, we consider the data-generating process in equations \eqref{2.1} and \eqref{2.2}. We set $\rho=0.16$, $\gamma = 0.15$, $\delta = 0.2$ and $\beta = -2$. These parameter values are chosen to ensure that the data-generating process mimics the findings from the first empirical application in the next section. 

We consider one exogenous variable drawn independently from the uniform distribution on the unit interval. We generate the error terms $\varepsilon_{t}(\xvec{s}_i)$'s independently from the standard normal distribution. The number of factors $q$ is set to two, and they are drawn independently from the standard normal distribution along with their loadings. The number of spatial units is set to $49$, and they are randomly assigned to a regular rectangular lattice. We consider a queen contiguity spatial weight matrix so that $m_{ij}$ is set to 1 if the $j$-th entity is adjacent to or shares a border with the $i$-th entity. Otherwise, $m_{ij}$ is set to zero. We then row normalize the weights matrix. The number of time periods takes values from $\{50, 100\}$. The length of the Markov chain is $100000$ draws, and the first $20000$ draws are discarded to dissipate the effects of the initial draws. 

We first estimate the dynamic spatiotemporal ARCH model with two factors using the Gibbs samplers in Algorithms 1 (standard MCMC) and 2 (shrinkage MCMC). We also consider the performance of the Gibbs samplers when the number of factors chosen by the researcher exceeds the true number of factors. To this end, we estimate the dynamic spatiotemporal ARCH model with three factors using the Gibbs samplers in Algorithms 1 and 2. Ideally, in this case, the Bayesian shrinkage estimator in Algorithm 2 shall shrink one of the factor loadings toward zero and shall perform better than the estimator in Algorithm 1.


\begin{table}[h] 
\centering 
\caption{\normalsize Simulation results}\label{tab-sim} 
\setlength{\tabcolsep}{3pt} 
{\footnotesize
\renewcommand{\arraystretch}{1.4} 
\begin{tabular}{l l c l l *{6}{c}} 
\toprule\toprule
$T$ & $q_{\max}$ & Shrinkage & & &$\rho$ & $\gamma$ & $\delta$ & $\beta$ & $\mathbf{h}^{*}$ & DIC \\ 
\hline 
\multirow{8}{*}{50} & \multirow{4}{*}{2} & \multirow{2}{*}{No (Alg. 1)} & median & &0.140&0.139&0.228&-1.992&-1.336&\multirow{2}{*}{7676.756} \\ 
                    &&                        &95\% CI& &[0.094,0.185]&[0.113,0.165]&[0.182,0.274]&[-2.102,-1.879]&[-4.356,1.683]& \\ 
                    &&\multirow{2}{*}{Yes (Alg. 2)} & median & &0.136&0.139&0.228&-1.972&-1.319&\multirow{2}{*}{7683.741} \\ 
                    &&                       &95\% CI & &[0.091,0.178]&[0.113,0.164]&[0.182,0.273]&[-2.083,-1.863]&[-4.029,1.457]& \\ 
\cline{2-11} 
& \multirow{4}{*}{3} & \multirow{2}{*}{No (Alg. 1)} & median & &0.140&0.138&0.235&-1.997&-1.361&\multirow{2}{*}{7666.523} \\ 
                    &&                        &95\% CI & &[0.092,0.186]&[0.110,0.166]&[0.187,0.283]&[-2.115,-1.878]&[-4.350,1.631]&  \\ 
                    &&\multirow{2}{*}{Yes (Alg. 2)}  & median & &0.137&0.138&0.233&-1.976&-1.337&\multirow{2}{*}{7737.805} \\ 
                    &&                        &95\% CI & &[0.093,0.179]&[0.112,0.164]&[0.187,0.278]&[-2.090,-1.863]&[-4.062,1.423]& \\ 
\midrule
\multirow{8}{*}{100} &\multirow{4}{*}{2} & \multirow{2}{*}{No (Alg. 1)} & median & &0.151&0.144&0.215&-2.017&-1.358&\multirow{2}{*}{15261.821} \\ 
                   &&                    &95\% CI & &[0.123,0.180]&[0.128,0.161]&[0.185,0.244]&[-2.089,-1.946]&[-4.835,2.085]& \\ 
                   &&\multirow{2}{*}{Yes (Alg. 2)} & median & &0.145&0.146&0.215&-2.015&-1.344&\multirow{2}{*}{15326.282} \\ 
                   &&                   &95\% CI & &[0.116,0.173]&[0.130,0.162]&[0.185,0.244]&[-2.086,-1.943]&[-4.574,1.839]& \\ 
\cline{2-11} 
&\multirow{4}{*}{3} & \multirow{2}{*}{No (Alg. 1)} & median & &0.161&0.147&0.213&-2.010&-1.385&\multirow{2}{*}{15262.991} \\ 
                   &&                    &95\% CI & &[0.131,0.191]&[0.129,0.163]&[0.182,0.243]&[-2.086,-1.934]&[-4.880,2.054]& \\ 
                   &&\multirow{2}{*}{Yes (Alg. 2)} & median & &0.149&0.148&0.217&-2.008&-1.367&\multirow{2}{*}{15363.761} \\ 
                   &&                   &95\% CI & &[0.120,0.177]&[0.131,0.164]&[0.187,0.247]&[-2.082,-1.934]&[-4.642,1.844]& \\ 
\bottomrule\bottomrule
\end{tabular} 
} 
\end{table} 


\begin{figure}[ht!]
    \centering  
     \includegraphics[width=3.3in, height=2.3in]{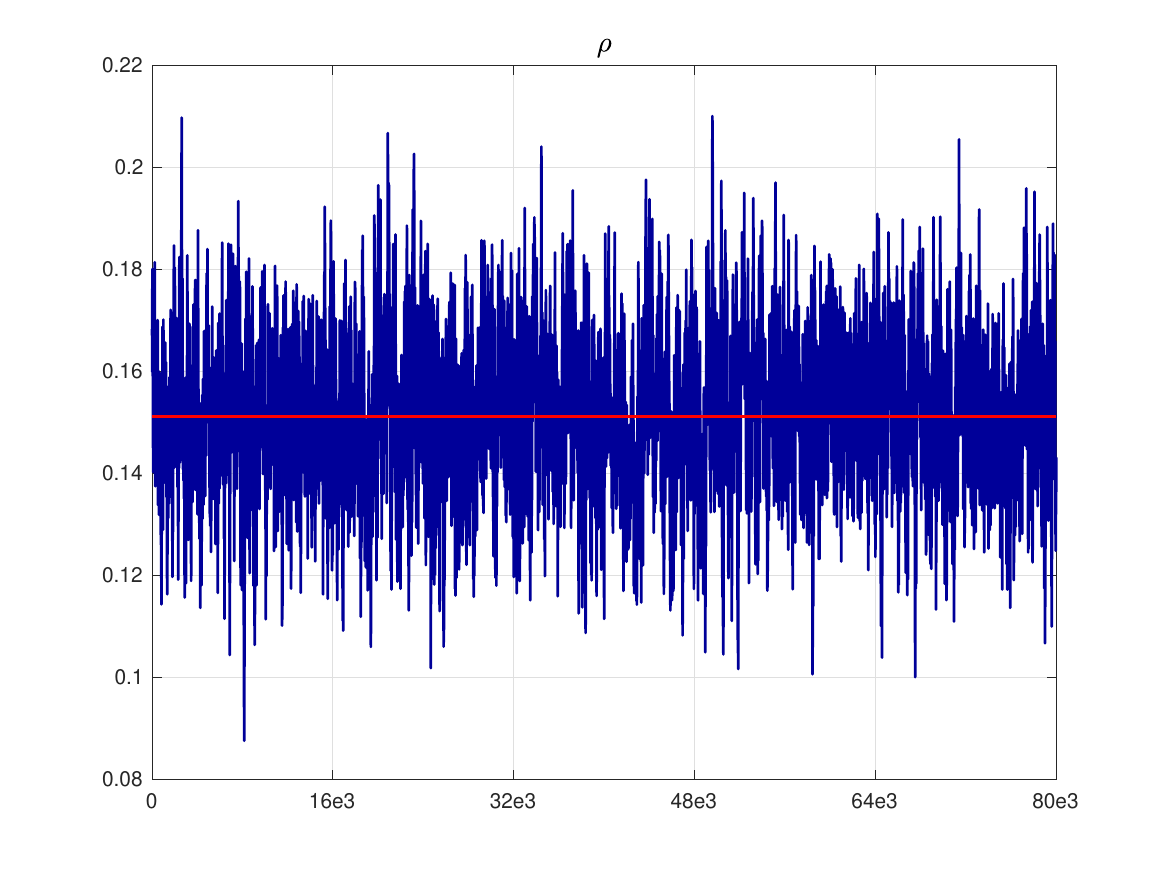}
     \hspace{-.89cm}
      \includegraphics[width=3.3in, height=2.3in]{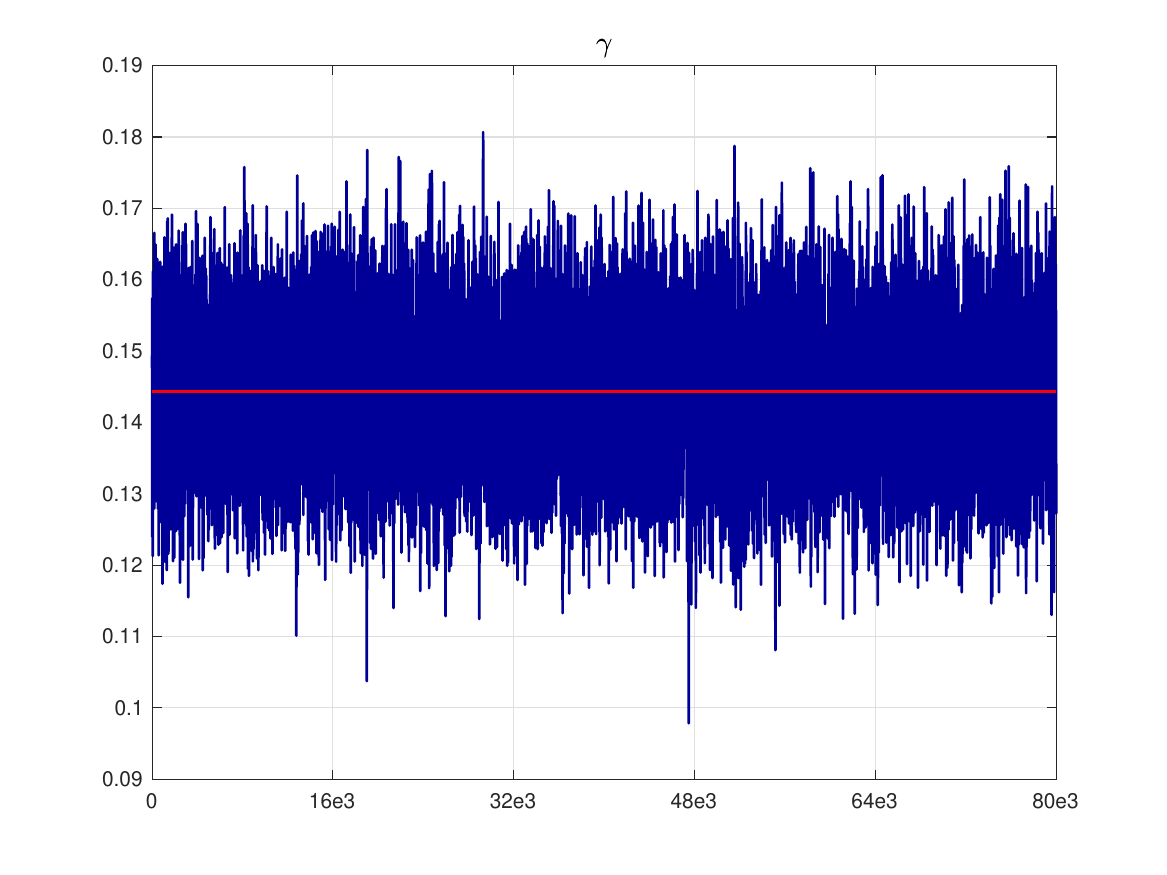}
      \includegraphics[width=3.3in, height=2.3in]{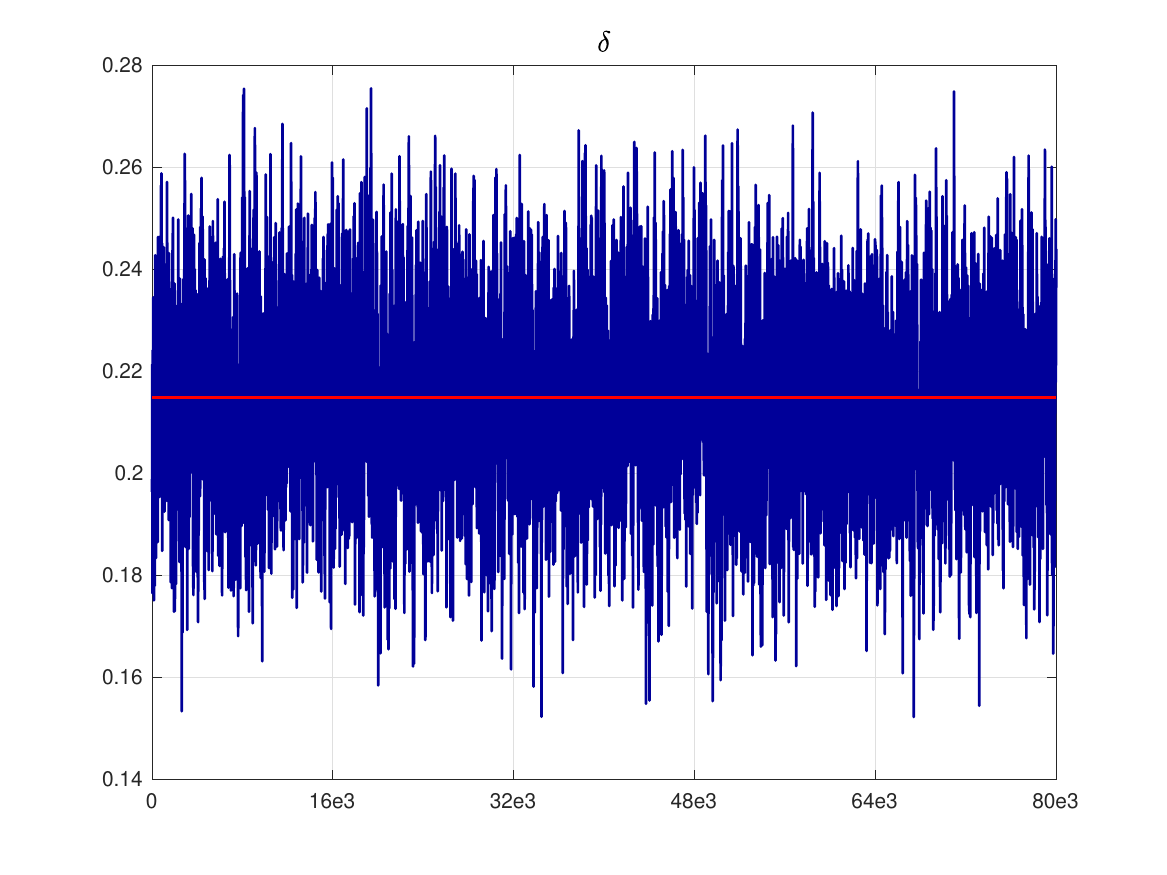}
     \hspace{-.89cm}
       \includegraphics[width=3.3in, height=2.3in]{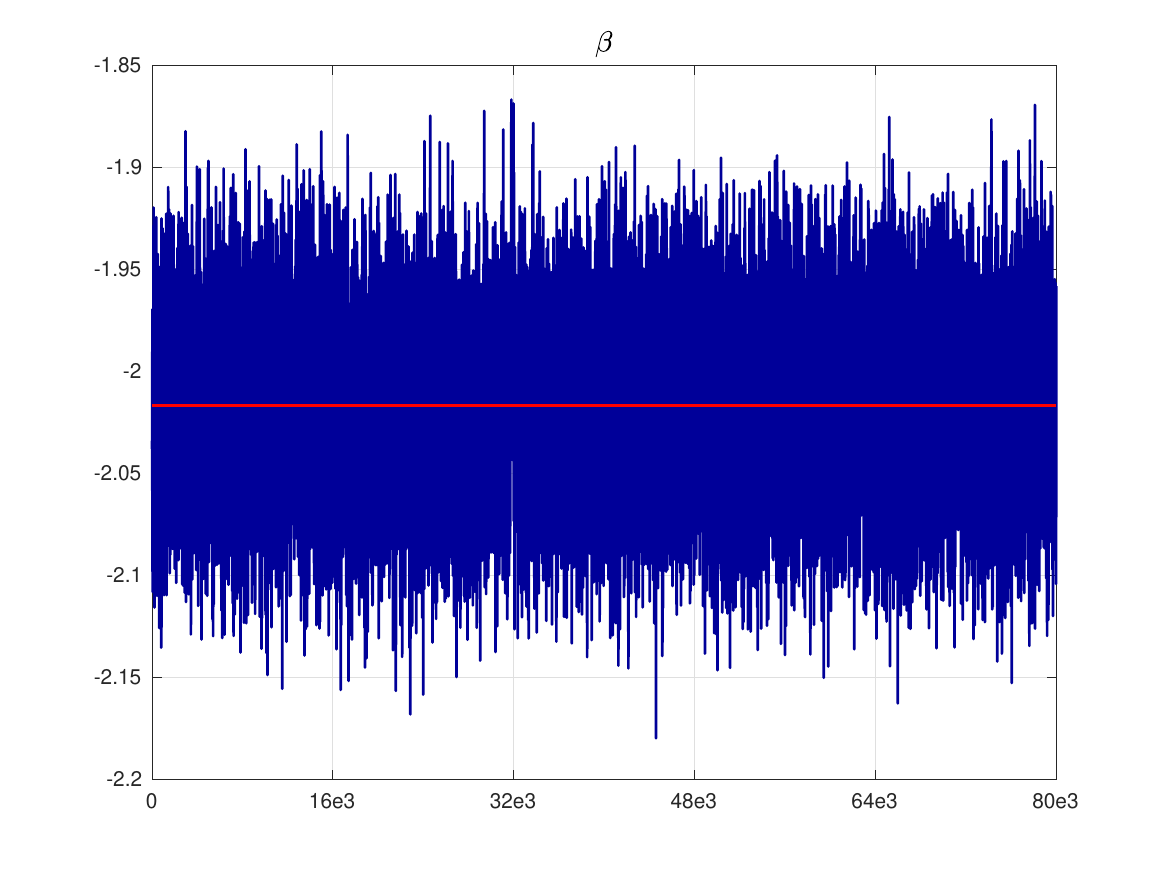}
     \caption{Trace plots for $\rho$, $\gamma$, $\delta$ and $\beta$: $T=100$, Algorithm 1, $q_{\max}=2$.}
\label{fig-sim1}		
\end{figure}

\begin{figure}[ht!]
    \centering  
     \includegraphics[width=3.3in, height=2.3in]{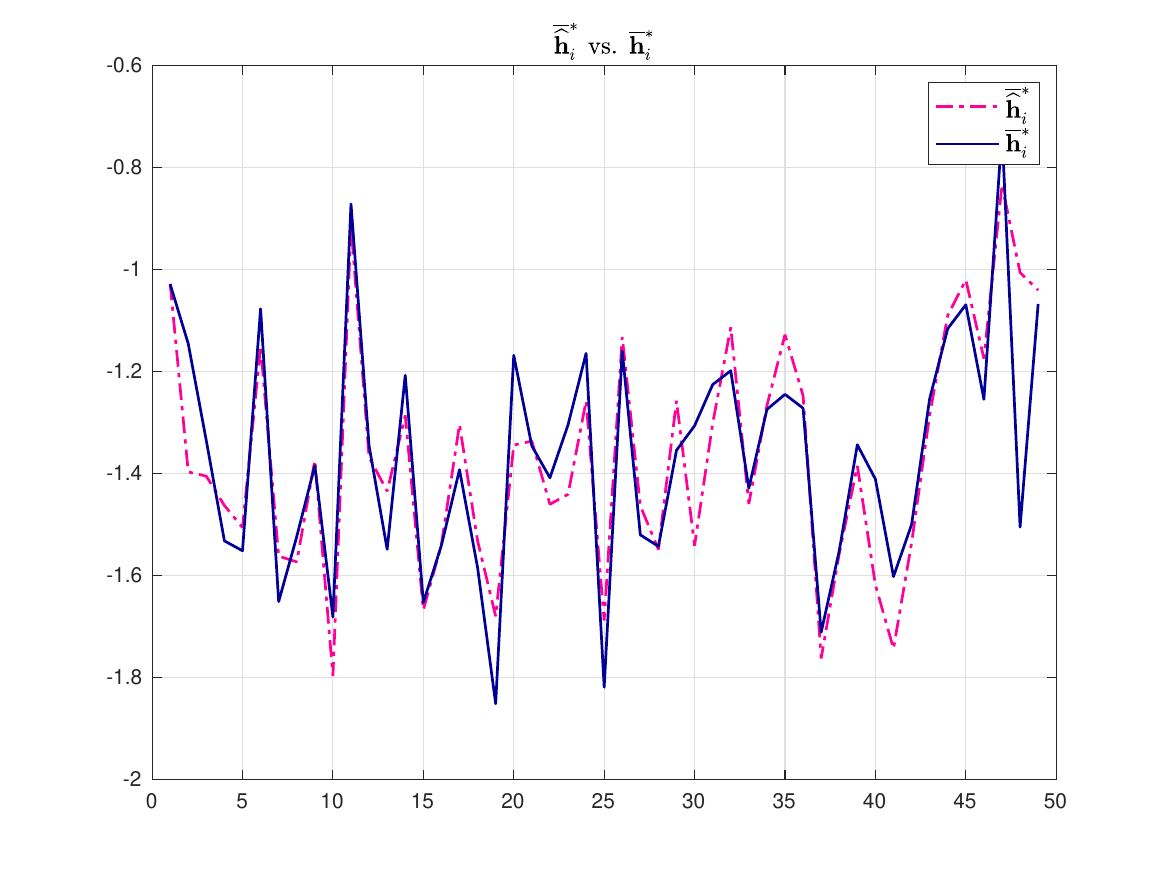}
     \hspace{-.89cm}
      \includegraphics[width=3.3in, height=2.3in]{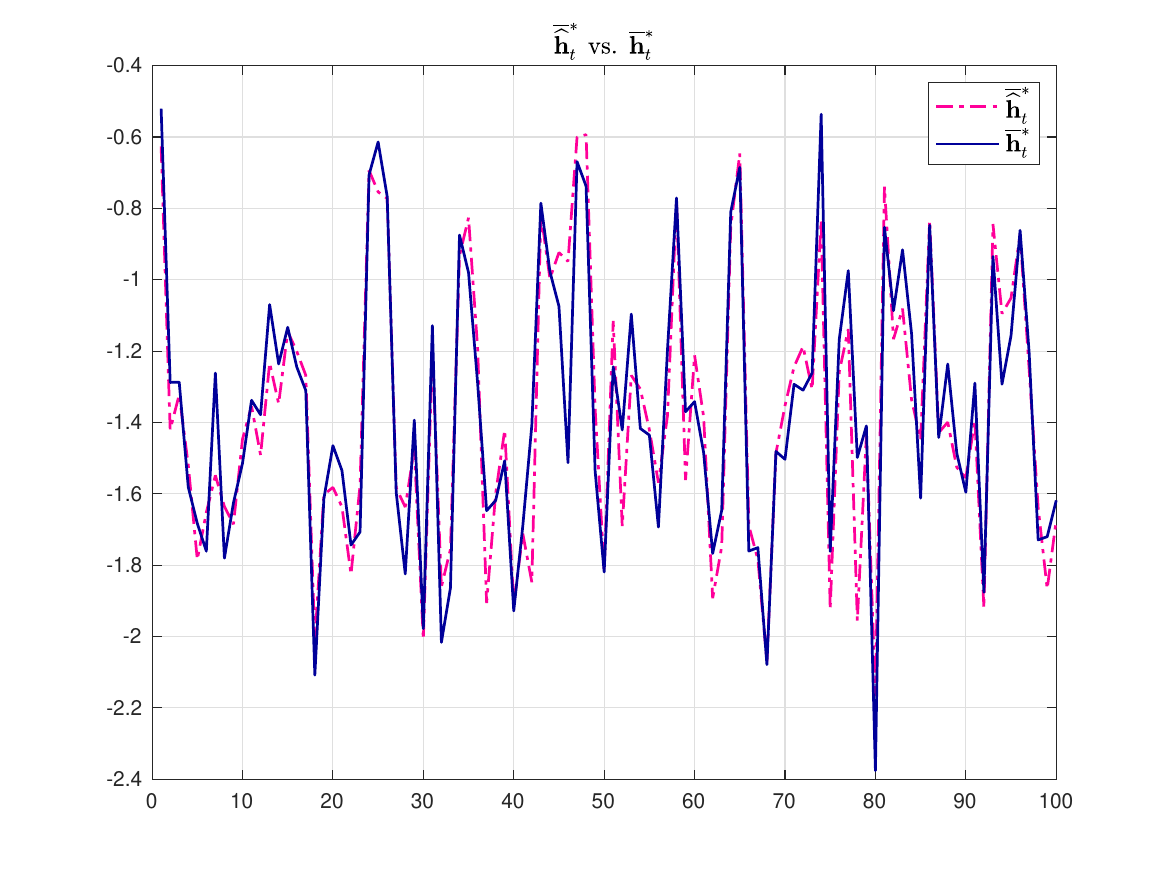}
     \caption{$\overline{\widehat{\mathbf{h}}}_i^*$ vs. $\overline{\mathbf{h}}_i^*$ and $\overline{\widehat{\mathbf{h}}}_t^*$ vs. $\overline{\mathbf{h}}_t^*$: $T=100$, Algorithm 1, $q_{\max}=2$.}
\label{fig-sim2}		
\end{figure}

Table~\ref{tab-sim} presents the results from the simulation exercise. We present the estimated posterior median and the 95\% credible interval for the parameters of the spatiotemporal ARCH model. We observe that both estimation algorithms report posterior median estimates that are close to the true values. The 95\% credible interval estimates indicate evidence for the precision of the suggested estimation algorithms. For $\mathbf{h}^{*}$, we report the overall average along with the 95\% credible interval using the median posterior draws. The true overall average value $\mathbf{h}^{*}$ is $-1.346$ when $T=50$ and $-1.354$ when $T=100$. The overall average estimates reported by both algorithms are close to the true averages. The DIC estimates indicate that when $T$ is short relative to $n$, the DIC may prefer the specification with a higher number of factors. However, when $T$ is large relative to $n$, the DIC chooses the model with the true number of factors.

To determine the adequacy of the length of the chains and their mixing properties, some exemplary trace plots are provided in Figure~\ref{fig-sim1}. For the sake of brevity, we only present the results for Algorithm 1 when $q_{\max}=2$ and $T=100$. In these trace plots, the red solid lines correspond to the estimated posterior means. We observe that the Bayesian sampler performs satisfactorily and mixes well in all cases. For $\mathbf{h}^{*}$, we calculate the average of true values over $n$ and over $T$, respectively, and plot them against the average posterior means over $n$ and over $T$, respectively. In Figure~\ref{fig-sim2}, we observe that the Bayesian sampler performs satisfactorily in terms of capturing the log volatility over spatial units as well as over time. 

\section{Applications with real data}\label{sec5}

\noindent In this section, we use our model to estimate volatility in the US house price returns at the state level and the US stock market returns. In addition to volatility estimates, our approach allows us to estimate the spatial, temporal, and spatiotemporal effects of log-squared returns on the log volatility of housing and stock markets.

\subsection{Modelling the house price volatility in the US}

For the first application, we consider the housing market in the US, which provides a natural setting to apply a spatio-temporal model \citep[e.g.,][]{baltagi2014further, holly2010spatio, valentini2013modeling}. Indeed, house prices are affected by spillovers from neighboring units, as houses share many common amenities. For example, \cite{kuethe2011regional}, examining housing prices in the western US states from 1998 to 2007, finds that in most states, prices are influenced by spatial spillovers. Analyzing data from 1975 to 2011, \cite{brady2014spatial} finds similar evidence across all US states. Similar findings are common for other housing markets outside the US, such as Germany \citep{otto2018spatiotemporal}, the Netherlands \citep{van2011modelling}, and the UK \citep{blatt2023changepoint}, among others. Moreover, it is well known that house prices are influenced by common factors originating from economy-wide shocks, such as changes in monetary policy, energy prices, and technology. Previous studies have shown that models incorporating common factors are better suited for modeling and predicting house prices compared to standard models \citep[e.g.,][]{bork2018housing, yang2021common}.

Spatial spillovers are also present in the volatility of house prices, a common measure of market risk. To account for this, \cite{tacspinar2021bayesian} suggest using a stochastic volatility model with spatial interactions for modeling house price volatility, while \cite{dougan2023bayesian, otto2023general, otto2024multivariate} propose spatiotemporal GARCH-type models, among others. While the presence of latent factors is crucial for explaining housing market volatility \citep{fairchild2015understanding}, models that incorporate both spatial spillovers and latent factors in volatility modeling remain largely overlooked. We contribute to the literature on housing market volatility with our proposed specification, which accounts for the spatial, temporal, and spatiotemporal effects of log-squared returns on log volatility while also allowing for the influence of common latent factors.

\begin{figure}[!htb]
 \includegraphics[width=0.99\textwidth]{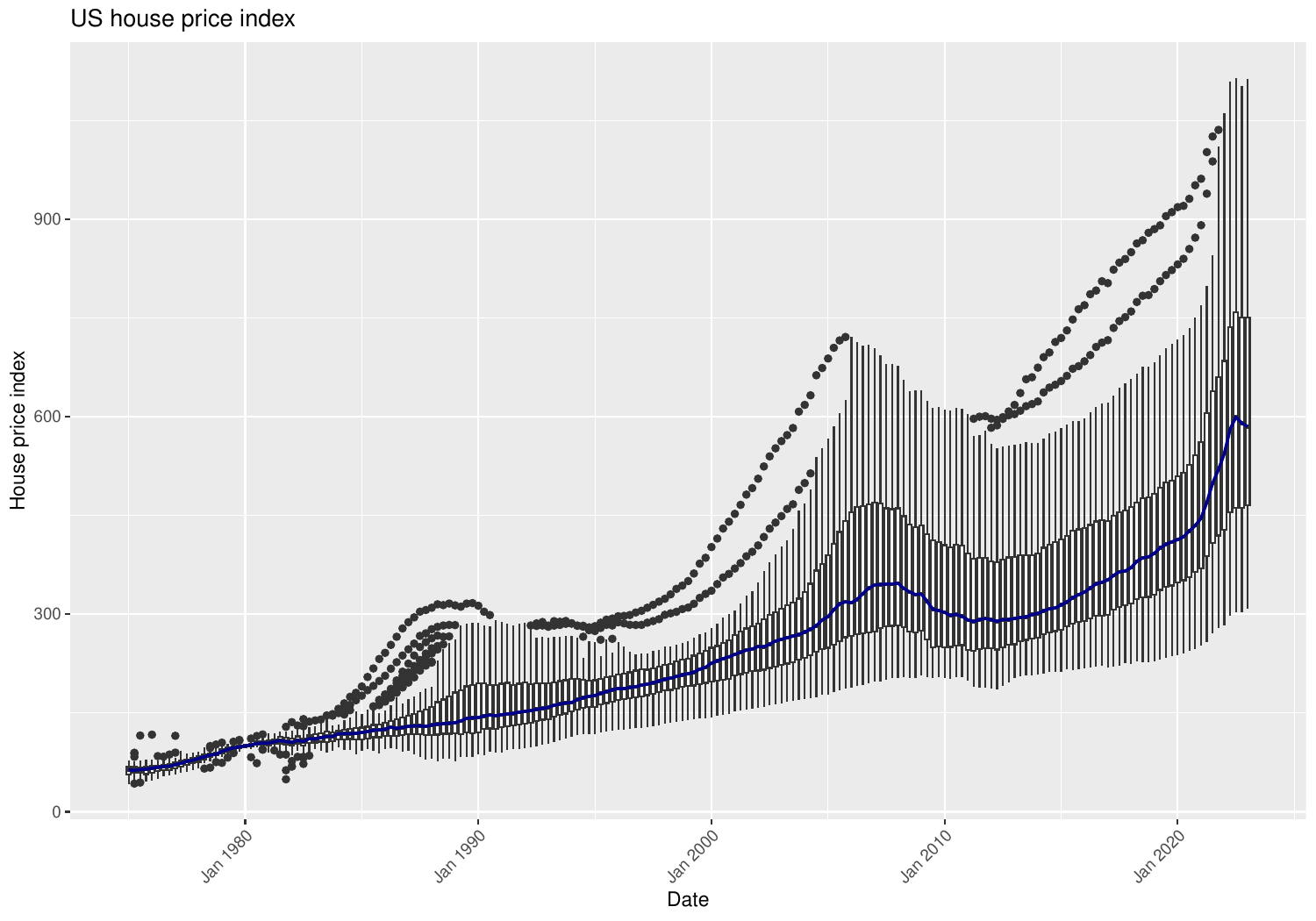}
	\vspace{-0.80cm}
	\caption{House price index time series for US states. Each boxplot represents the price distribution at a given point in time, and the solid line indicates the median price across all US states.}
 \label{fig:fig4}
	\end{figure}

For this application, we consider the quarterly house price indexes for the continental US states between 1975 and 2023. The data source is the Federal Housing Finance Agency website\footnote{Data can be retrieved from the following link: \url{https://www.fhfa.gov/}}. Our dataset includes $N=49$ spatial units and $T=193$ quarterly observations. We incorporate the average house price index for the US as an observable factor in the model. Figure~\ref{fig:fig4} shows the temporal evolution of the house price index for the US, where each boxplot represents the distribution of prices at a given point in time across all US states. We observe a generally positive trend with some local peaks, while the spread of the index across states also increases over time. Additionally, the index distribution appears to be right-skewed, with an increasing degree of asymmetry over time.

To measure spatial interaction in volatility dynamics, following previous studies, we consider a row-standardized contiguity spatial weight matrix $\mathbf{M}$. That is, all directly neighboring states are equally weighted, and only direct neighbors influence the price index in a given state. To select the number of latent factors in our specification, we use the DIC and Bayesian Lasso approaches presented in Section 3. We compute the DIC criterion for different values o $q$, using both Algorithms 1 and 2. The results are presented in Table~\ref{tab:dic_house1}. Eventually, we choose the $q$ value that yields the lowest DIC criterion, specifically selecting the model with $q=7$ latent factors. 

\begin{table}[!htb]
\centering
\begin{tabular}{lc}
\toprule
 & \textbf{DIC} \\
\midrule
\multicolumn{2}{l}{Algorithm \ref{a1}}\\
\midrule
$q=1$ & 27952.94 \\
$q=2$ & 27628.69 \\
$q=3$ & 27563.84 \\
$q=4$ & 27482.38\\
$q=5$ & 27444.69\\
$q=6$ & 27444.74\\
$q=7$ & \textbf{27415.85} \\
$q=8$ & 27435.50\\
\midrule
\multicolumn{2}{l}{Algorithm \ref{a2}}\\
\midrule
$q_{\max}=8$ & 28055.36 \\
\bottomrule
\end{tabular}
\caption{DIC values for Network log-ARCH models with varying numbers of common factors $q$ -- housing market dataset and contiguity-based spatial weight matrix. Both Algorithm \ref{a1} and Algorithm \ref{a2} are considered. The best model, highlighted in bold, is the one with the lowest DIC value.}
\label{tab:dic_house1}
\end{table}

House price volatility can be driven by several economy-wide underlying factors of various types. For instance, we may encounter several macroeconomic factors influenced by global economic conditions, such as interest rates, employment, energy prices, wage growth, and technology. However, supply and demand also play a crucial role; factors like rising construction costs or population decline can either increase or depress prices. Investor activity is another important factor that can influence housing markets. For example, changes in investor sentiment—often triggered by speculative bubbles bursting—can lead to sharp price corrections, contributing to market volatility. Therefore, incorporating a larger number of factors allows us to account for the effects of many of these influences.

Using Algorithms~\ref{a1} and \ref{a2}, we estimate our specification by setting the length of the MCMC chains to $100000$, discarding the first $20000$ draws as burn-ins. The trace plots for the parameters $\beta$, $\gamma$, $\rho$, and $\delta$ show stable behavior, indicating convergence of the algorithm. The median posterior draws serve as reliable point estimates for the model parameters. Recall that $\beta$ is the parameter for the observable factor (representing market volatility), $\gamma$ denotes the temporal effect parameter, $\rho$ is the spatial effect parameter, and $\delta$ is the spatiotemporal effect parameter.


We present detailed estimation results based on Algorithms~\ref{a1} and \ref{a2} in Table~\ref{tab:estimates_house1}. In this table, we provide the estimated posterior median for each parameter, $\bs{\lambda}^{'}_i \bs{f}_t$, and $\log h_t(\mf(s)_i)$ along with their 95\% credible intervals. We include estimation results for $q$ ranging from $1$ to $8$ and highlight the selected model by the DIC criterion in gray. In this table, statistically significant coefficients are those whose confidence intervals do not include zero.

\begin{table}[!htb]
    \centering
   \scalebox{0.8}{\begin{tabular}{lcccc|cc}
        \toprule
Factors   & $\beta$ & $\gamma$ & $\delta$ & $\rho$ & $\boldsymbol{\lambda}'\boldsymbol{f}_t$ & $\log h_t(\mathbf{s}_i)$ \\
        \midrule
        \multicolumn{7}{l}{Panel A: Algorithm \ref{a1}, DIC selection}\\
        
 $q=1$ & -2.8796 & 0.1645 & 0.1409 & 0.1621 & -3.8525 & -8.2637 \\
           & (-5.1635, -0.5912) & (0.1515, 0.1775) & (0.1165, 0.1660) & (0.1437, 0.1810) & (-5.2063, -1.7108) & (-9.7058, -5.1585) \\
$q=2$ & -2.5434 & 0.1502 & 0.1667 & 0.1811 & -3.6707 & -8.3101 \\
           & (-4.9368, -0.1247) & (0.1351, 0.1653) & (0.1399, 0.1915) & (0.1592, 0.2036) & (-4.8933, -1.8849) & (-9.6470, -5.4370) \\
 $q=3$  & -2.2541 & 0.1392 & 0.1732 & 0.1834 & -3.7155 & -8.3090 \\
           & (-4.6826, 0.1861) & (0.1232, 0.1553) & (0.1455, 0.2000) & (0.1566, 0.2031) & (-4.8906, -1.9410) & (-9.6675, -5.4981) \\
 $q=4$ & -2.1066 & 0.1340 & 0.1791 & 0.1679 & -3.8639 & -8.3065 \\
           & (-4.4739, 0.3032) & (0.1170, 0.1508) & (0.1470, 0.2086) & (0.1404, 0.1911) & (-5.0725, -2.0790) & (-9.7616, -5.5348) \\
 $q=5$  & -2.1026 & 0.1313 & 0.1831 & 0.1607 & -3.9073 & -8.3015 \\
           & (-4.5737, 0.4427) & (0.1142, 0.1481) & (0.1503, 0.2140) & (0.1372, 0.1958) & (-5.1147, -2.1218) & (-9.7689, -5.5680) \\
 $q=6$  & -2.1567 & 0.1356 & 0.1867 & 0.1650 & -3.7910 & -8.2876 \\
           & (-4.6468, 0.2916) & (0.1176, 0.1539) & (0.1502, 0.2205) & (0.1349, 0.1950) & (-5.0102, -2.0614) & (-9.7572, -5.6072) \\
\rowcolor{lightgray}
 $q=7$  & -2.0728 & 0.1461 & 0.1911 & 0.1609 & -3.6715 & -8.2494 \\
\rowcolor{lightgray}
           & (-4.5731, 0.4229) & (0.1278, 0.1645) & (0.1611, 0.2246) & (0.1342, 0.1858) & (-4.9100, -1.9667) & (-9.7521, -5.6340) \\
 $q=8$  & -2.0261 & 0.1480 & 0.1911 & 0.1600 & -3.6893 & -8.2502 \\
           & (-4.5976, 0.5365) & (0.1292, 0.1665) & (0.1592, 0.2216) & (0.1290, 0.1848) & (-4.9085, -1.9777) & (-9.7530, -5.6381) \\
         \midrule
 \multicolumn{7}{l}{Panel B: Algorithm \ref{a2}, Bayesian LASSO}\\
    $q_{\max}=8$ & -4.1026 & 0.1950 & 0.3429 & 0.3190 & -0.3917 & -8.2200 \\
           & (-6.7235, -1.5565) & (0.1769, 0.2130) & (0.3183, 0.3696) & (0.2894, 0.3424) & (-1.3351, 0.6861) & (-9.6560, -5.5954) \\
        \bottomrule
    \end{tabular}}
    \caption{Median posterior estimates with lower and upper bounds (0.025, 0.975) of the Network log-ARCH with common factors -- housing market data with contiguity-based spatial weighting matrix. Different numbers of factors $q$ are in the rows and the parameters are in the column. The last row shows the results obtained from Algorithm~\ref{a2}, assuming $q_{\max}=8$. Statistically significant coefficients are those with the confidence interval not including the zero. The selected model by the DIC criterion is $q = 7$, which is highlighted in gray.}
    \label{tab:estimates_house1}
\end{table}


Focusing on the results obtained with $q=7$, highlighted in gray, we find that all parameter estimates are statistically significant, except for the estimate of $\beta$. The estimate of $\rho$ indicates that the house price volatility of a state is positively affected by the log-squared house price returns of neighboring states. Thus, this parameter represents the contemporaneous spillover effect, underscoring how current volatility from the housing markets in the neighboring states influences the housing sector in real time. The positive and statistically significant estimate of the spatiotemporal effect ($\delta$) shows that past shocks in house markets in the neighboring states continue to affect housing volatility. In both cases, we observe statistically significant spillover effects in the US housing market. The estimates of $\rho$ and $\delta$ are $0.16$ and $0.19$, respectively. Therefore, both instantaneous and lagged spillovers have similar effects in terms of magnitude, with the spatiotemporal effect being slightly larger than the instantaneous spatial effect. The estimated temporal effect ($\gamma$) is positive and statistically significant across all models, suggesting that volatility persists over time within the housing market, with current volatility being influenced by past market risks. This persistence is key to understanding the cyclical nature of housing market volatility, where a shock today continues to have repercussions in the future. Finally, we find that the volatility of the US house price index $(\beta)$, obtained as the average of the state-specific house price indexes, is not statistically significant. This is probably because the market effect is better captured by some of the $q$ latent factors.

\begin{landscape}
\begin{figure}
	\begin{center}
	    \includegraphics[width=9in]{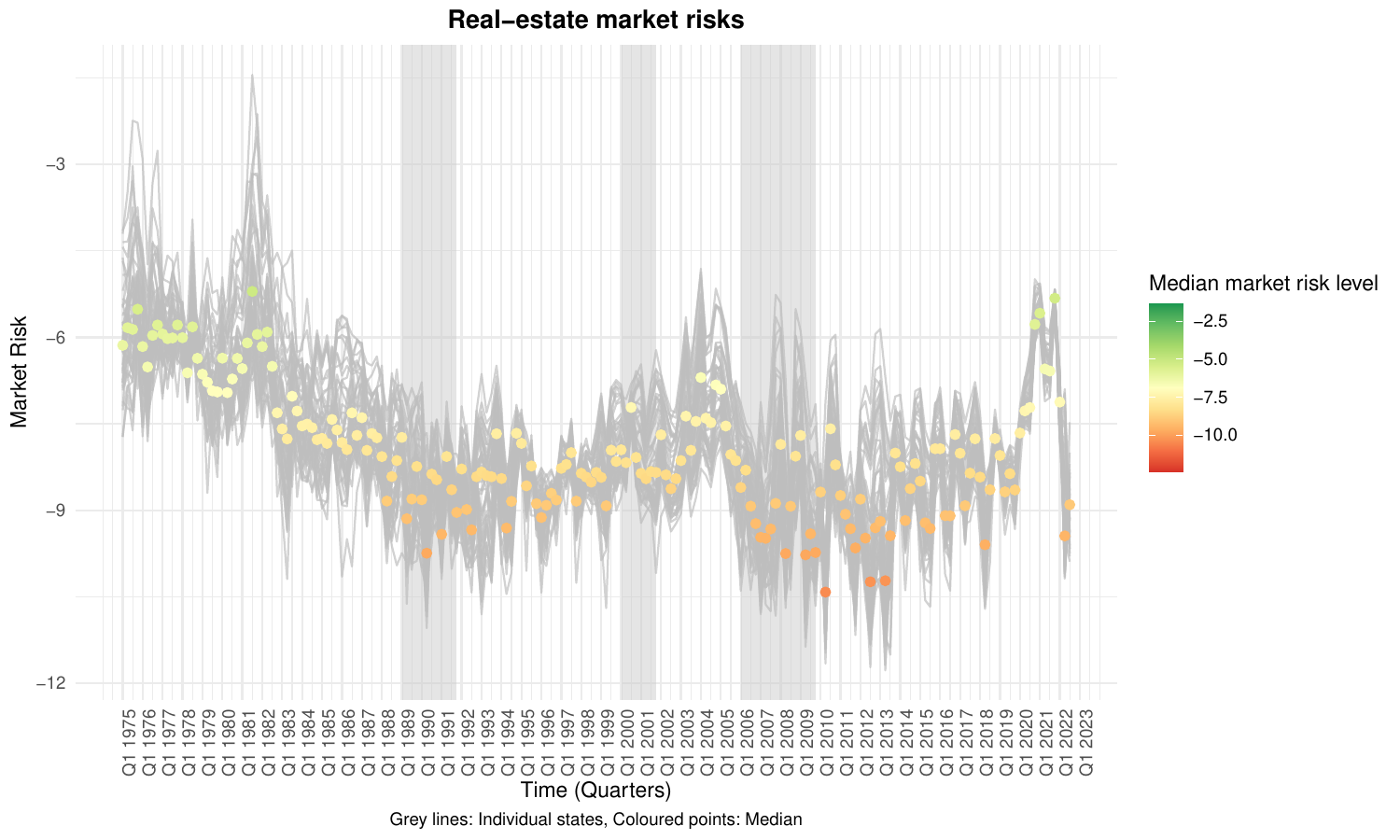}
	\end{center}
	\vspace{-0.80cm}
	\caption{Median posterior draws of the log volatilities $\widehat{\log h_{t}(\mf{s}_i)}$ plotted across time for all states based on $q=7$ factors. Each grey line corresponds to one state, while the colored points show the medians across all states. The three shaded time periods correspond to the early 1990s crisis, the dot bubble, and the great financial crisis. These periods are analyzed with more detail in Figures~\ref{fig:hstar_house_map1}-\ref{fig:hstar_house_map3}.}
	\label{fig:hstar_house_timeseries}
\end{figure}
\end{landscape}

Next, we analyze the estimated values of the log volatility in the US housing market. Figure~\ref{fig:hstar_house_timeseries} shows the median posterior draws of the log volatility $\widehat{\log h_{t}(\mf{s}_i)}$, plotted across time for all states based on $q=7$ latent factors. In the figure, each grey line corresponds to one state, while the colored points show the medians across all states. We highlight three different time periods in Figure~\ref{fig:hstar_house_timeseries}, each characterized by a shock period. The first one corresponds to the early 1990s recession in the US, between 1989 and 1991; the second one is the dot-com bubble between 2000 and 2001, while the third one corresponds to the great financial crisis of 2007--2009 and the period just before the crisis occurred.

Figure~\ref{fig:hstar_house_timeseries} shows that spikes of volatility are not always observed during economic crises. Between 1989 and 1990, the US economy was weakening due to a restrictive monetary policy, exacerbated by the 1990 oil shock. Despite these shocks, we do not observe a clear spike in the volatility of the housing market during this time period. The opposite applies, however, to the dot-com bubble in the early 2000s. As happened in the stock market, we also observed an increase in the volatility of housing prices during this time period. Finally, for the great financial crisis, we observe a significant spike in US house price volatility because volatility follows a decreasing pattern in the period immediately before the crisis, that is, between 2006 and early 2007. Interestingly, we find increasing volatility during COVID-19, that is, between 2020 and 2022, while we observe a strong decrease in US housing market volatility after the shock.

\begin{figure}[!htb]
	\begin{center}
	    \includegraphics[width=0.99\textwidth]{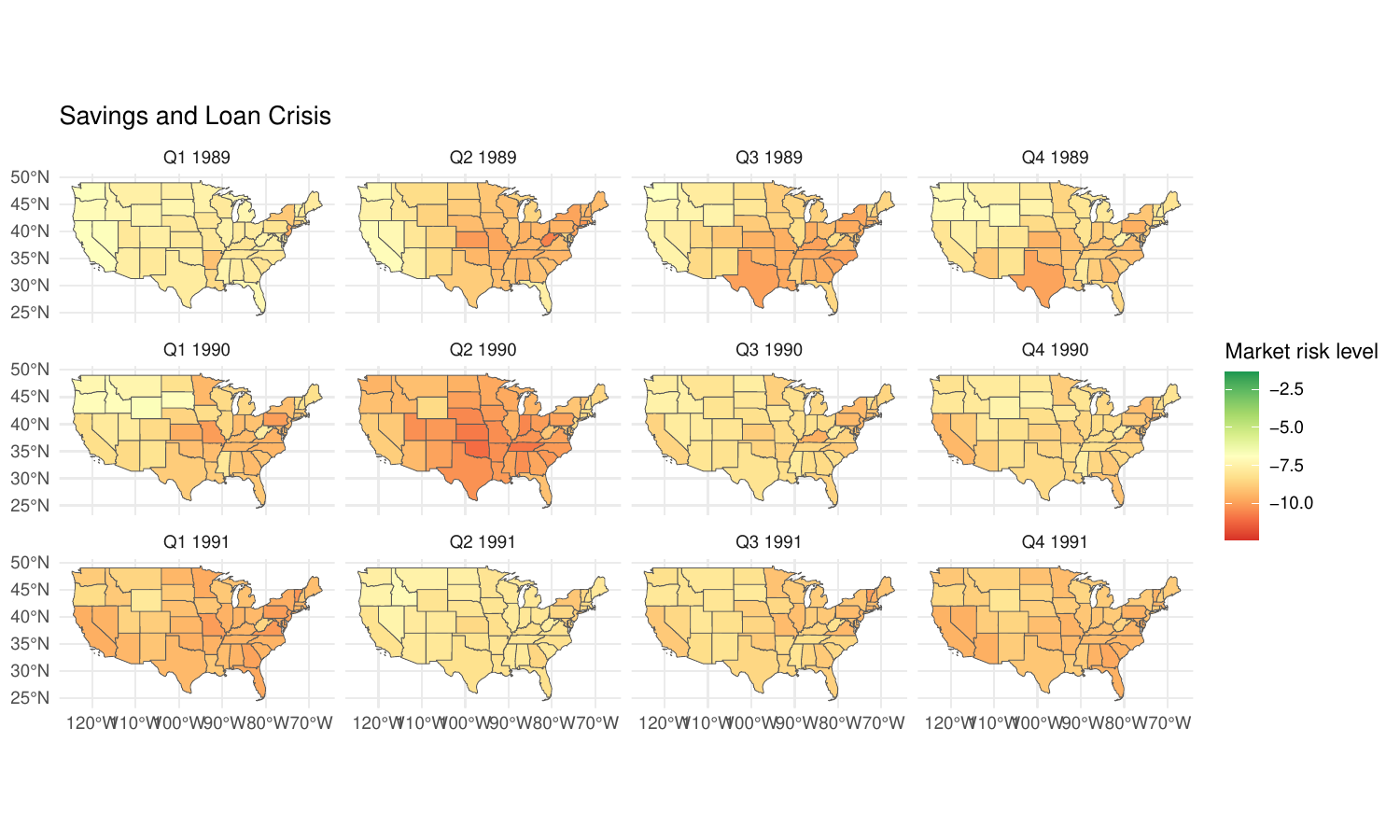}
	\end{center}
	\vspace{-0.80cm}
	\caption{Median posterior draws of the log volatilities $\widehat{\log h_{t}(\mf{s}_i)}$ from 1989 to 1991 (savings and loan crisis) plotted on the map (7 factors).}
	\label{fig:hstar_house_map1}
\end{figure}

\begin{figure}[!htb]
	\begin{center}
	    \includegraphics[width=0.99\textwidth]{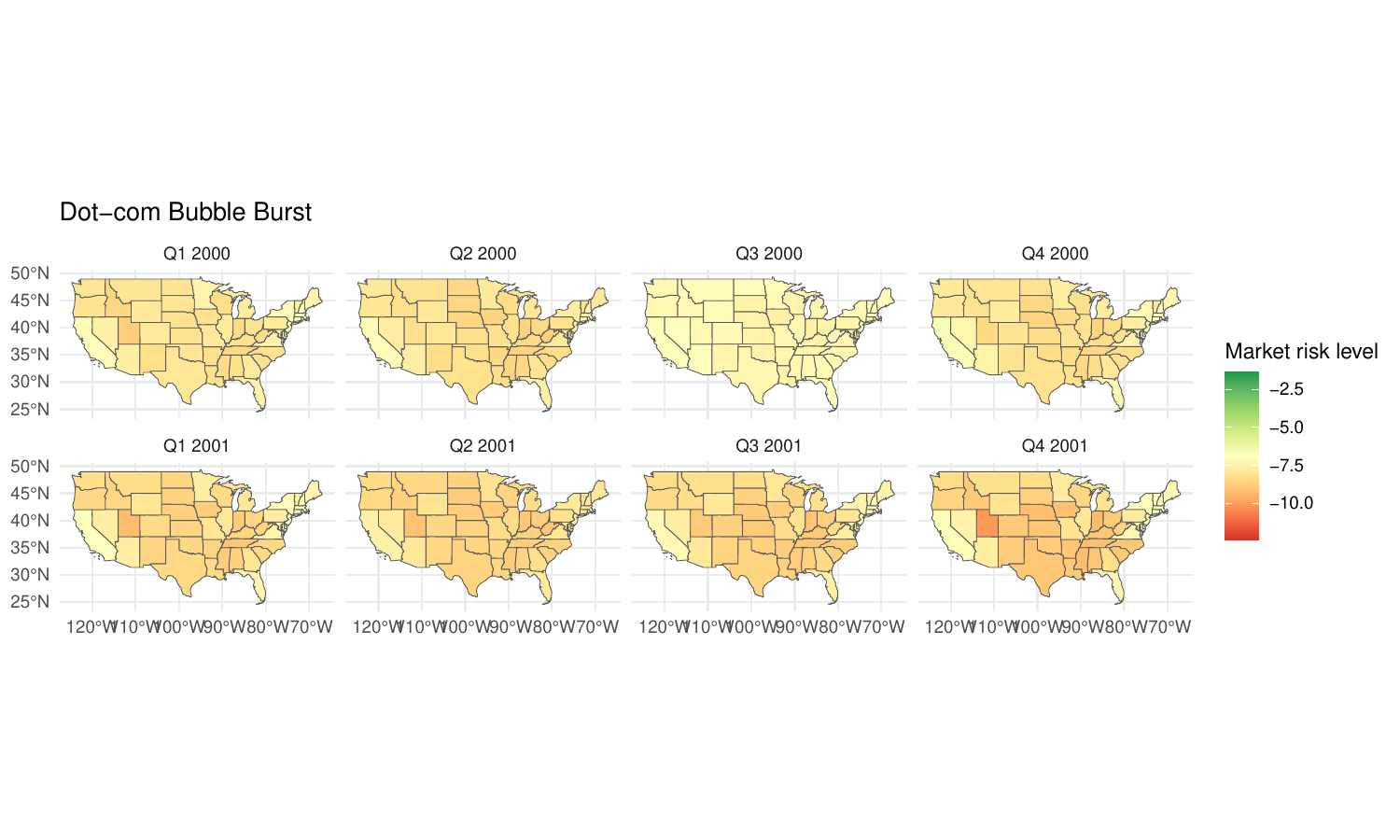}
	\end{center}
	\vspace{-0.80cm}
	\caption{Median posterior draws of the log volatilities $\widehat{\log h_{t}(\mf{s}_i)}$ from 2000 to 2001 (dot-com bubble) plotted on the map (7 factors).}
	\label{fig:hstar_house_map2}
\end{figure}

\begin{figure}[!htb]
	\begin{center}
	    \includegraphics[width=0.99\textwidth]{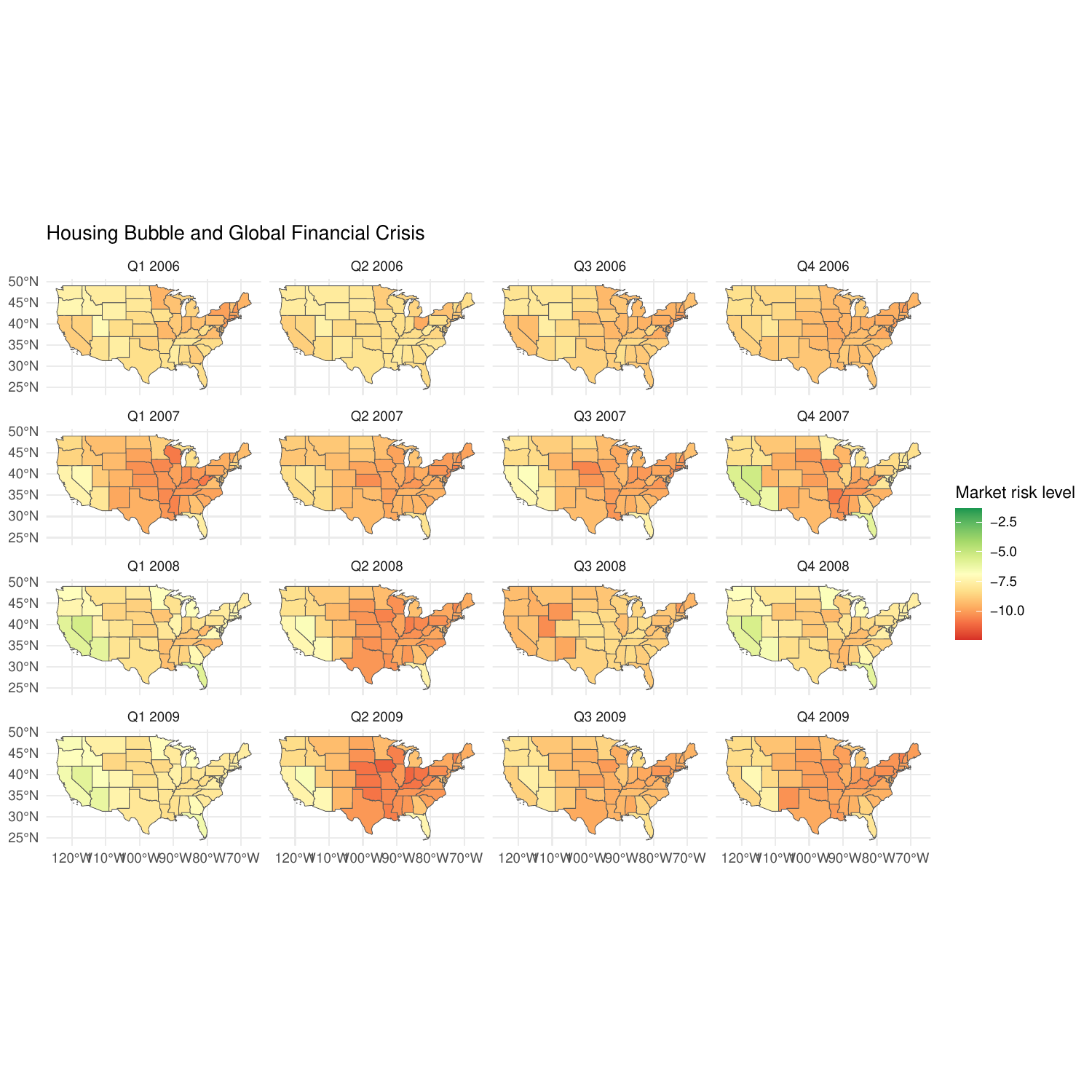}
	\end{center}
	\vspace{-0.80cm}
	\caption{Median posterior draws of the log volatilities $\widehat{\log h_{t}(\mf{s}_i)}$ from 2006 to 2009 (housing bubble and great financial crisis) plotted on the map (7 factors).}
	\label{fig:hstar_house_map3}
\end{figure}

In Figures~\ref{fig:hstar_house_map1}-\ref{fig:hstar_house_map3}, we provide a detailed analysis of the regional differences in the estimated volatilities during the three considered periods. In these figures, the median posterior draws of volatilities based on the selected model with $q=7$ are shown through the US states maps. The color scale in these figures corresponds to the color scale in Figure~\ref{fig:hstar_house_timeseries}.

Considering the savings and loan crisis between 1989 and 1991 (see Figure~\ref{fig:hstar_house_map1}), we find that states in the West are characterized by larger volatility compared to states in the East of the country. However, the states in the Northeast, such as New Hampshire, New York, and New Jersey, show relatively larger volatility. These results confirm the existence of positive spatial spillovers in the volatility dynamics. Figure~\ref{fig:hstar_house_map2} shows the spatial volatility patterns during the dot-com bubble. In this case, the spatial differences become even more evident, as states in the West show larger volatility compared to those in the center and in the East, with the exception of northeastern states that are characterized by high volatility during the whole period.

Finally, we consider the volatility patterns during the great financial crisis. The results are shown in Figure~\ref{fig:hstar_house_map3}. In the quarters before the crisis, that is, during 2006, we find that all the states are characterized by relatively low volatility levels. From the first quarter of 2007, we observe that southwestern states such as Arizona, California, and Nevada show the highest volatility in the entire country. Starting from the last quarter of 2007, however, Florida and the northeastern states also begin to show high volatility levels. Arizona, California, and Nevada, in particular, exhibit a considerable increase in volatility. The first and last quarters of 2008, as well as the first quarter of 2009, show the highest levels of volatility for the entire US. After this point, we observe a sudden decrease in volatility for the remaining quarters in Figure~\ref{fig:hstar_house_map3}.

\FloatBarrier

Overall, the detailed analysis of these periods shows peculiar spatial patterns; that is, western and northeastern states are characterized by large volatility, while central states of the US have lower volatility levels. Moreover, our model allows us to obtain individual volatility or real-estate market risk levels for each state and time point. 

\subsection{Modelling the volatility of the US stock market}

\begin{figure}[!htb]
	\begin{center}
	    \includegraphics[width=0.90\textwidth]{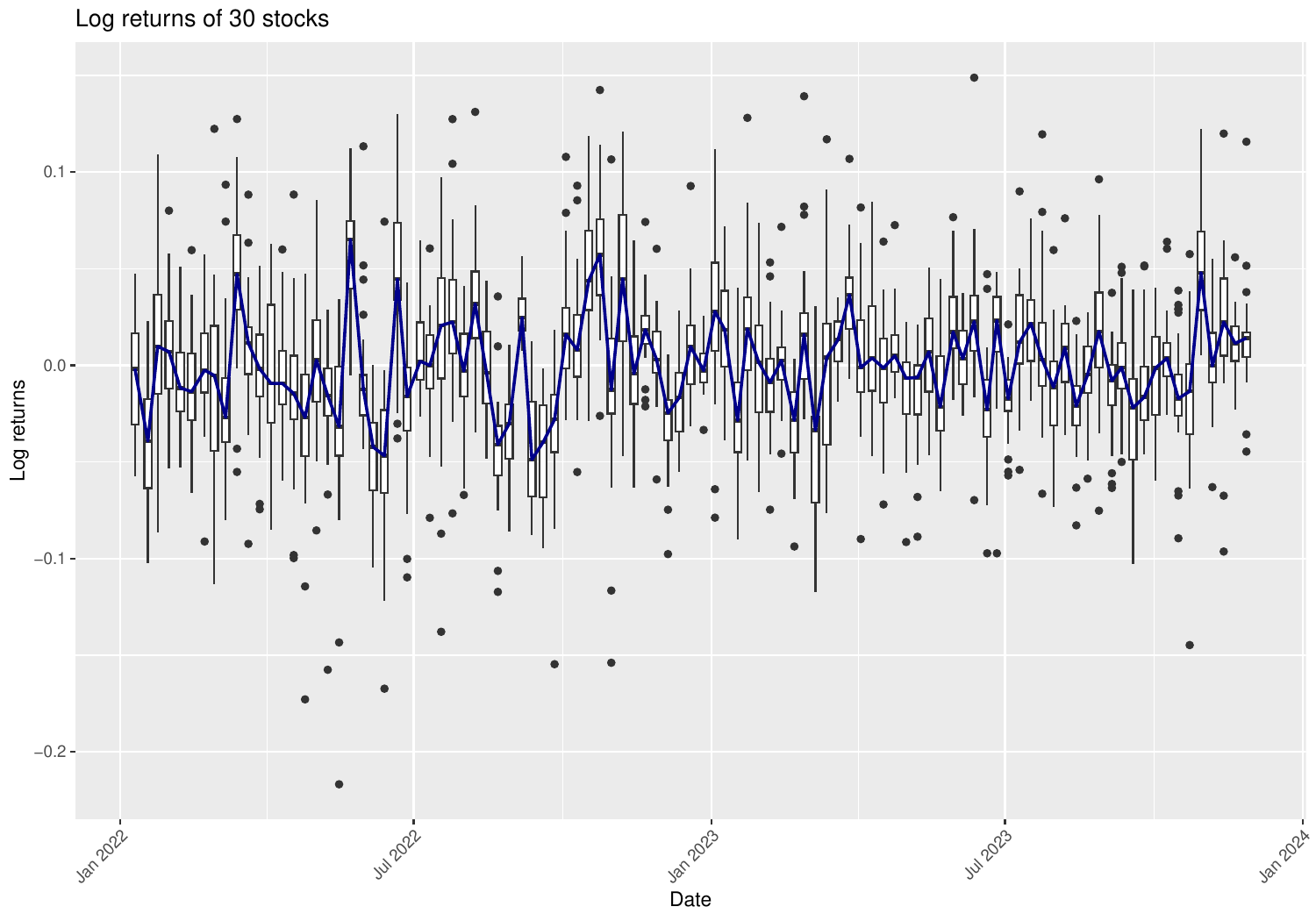}
     \includegraphics[width=0.90\textwidth]{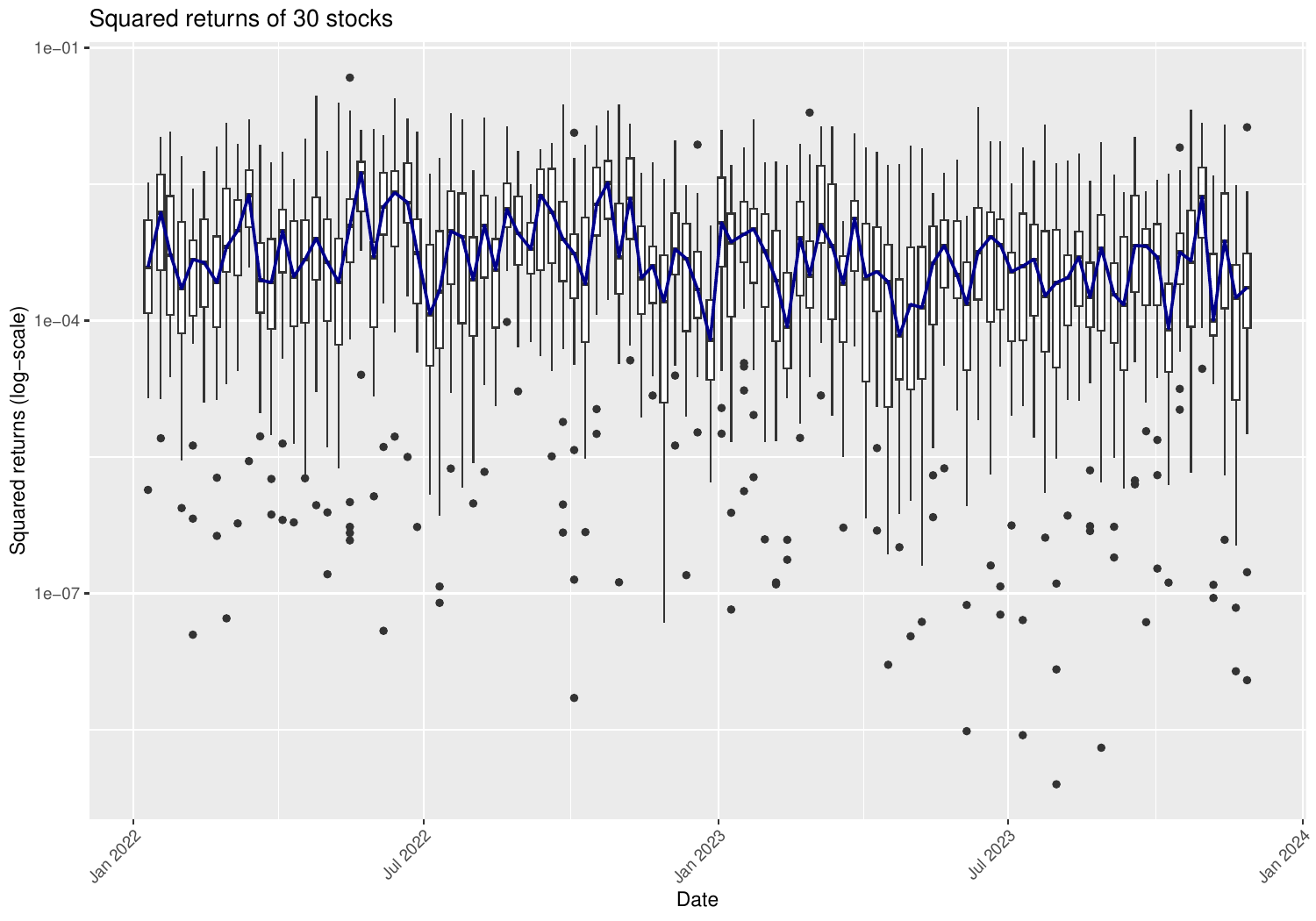}
	\end{center}
	\vspace{-0.80cm}
	\caption{Weekly log-returns (top) and squared log-returns (bottom) of the $N=30$ stocks included in the Dow Jones Industrial Average Index over the period 2022-2023. The weekly returns were summarised as boxplots at each time point. The solid line represents the median returns across all 30 stocks.}
	\label{fig:data_time}
\end{figure}

In this section, we consider an application of the proposed model for analyzing financial volatility in a stock market network. We study the weekly time series of the stocks included in the Dow Jones Industrial Average over the last two years. Therefore, we consider $N=30$ stocks over a time period with length $T=100$. Figure~\ref{fig:data_time} shows the weekly log returns (top figure) and squared log returns (bottom figure) of the $N=30$ stocks included in the Dow Jones Industrial Average Index during the period 2022-2023. The weekly returns are summarized as boxplots, showing the cross-sectional variation of the log returns and squared log returns at each point in time, while the solid line represents the median returns across all $N$ stocks.

Figure~\ref{fig:data_time} highlights an important cross-sectional variability in terms of returns temporal patterns, with many outliers stocks at different points in time. For better modeling the volatility patterns of these stocks, it may be useful to consider the information of similarly outlying stocks. Moreover, consistent with previous studies \citep{alizadeh2002range,hansen2006consistent}, Figure \ref{fig:data_time} also shows that the log squared returns time series is a very noisy proxy of volatility, as it is subject to significant random fluctuations and may not consistently reflect the underlying volatility patterns. We are, therefore, interested in getting a suitable estimate of the volatility in this application.  

Before presenting the estimation results, we compare our specification with commonly used GARCH models and their exponential extensions, such as log-ARCH models. Compared to the standard log-ARCH model, the proposed network log-ARCH approach provides two key features for volatility estimation. First, it accounts for contemporaneous and temporally lagged volatility spillovers through the parameters $\rho$ and $\delta$, which are well acknowledged in the financial literature \citep[e.g., see][]{diebold2014network,diebold2015trans}. Second, it considers the effect of latent economy-wide factors that influence volatility across the entire stock market \citep[e.g.,][]{han2006asset,barigozzi2016generalized,calzolari2021latent}. The inclusion of latent factors is particularly important when considering stocks traded in the same market index, as the effect of latent factors is stock-specific in the model. Since the model allows for the inclusion of observed factors, $\mathbf{x}_t(\mathbf{s}_i)$, we consider market volatility—the log-volatility of the Dow Jones Index—as an observed covariate in the model. As a result, our volatility estimation accounts for the spatial effect, the temporal effect, the spatiotemporal effect, the economy-wide latent factors effect, and the observed market volatility.

\begin{figure}[!htb]
	\begin{center}
	    \includegraphics[width=0.48\textwidth]{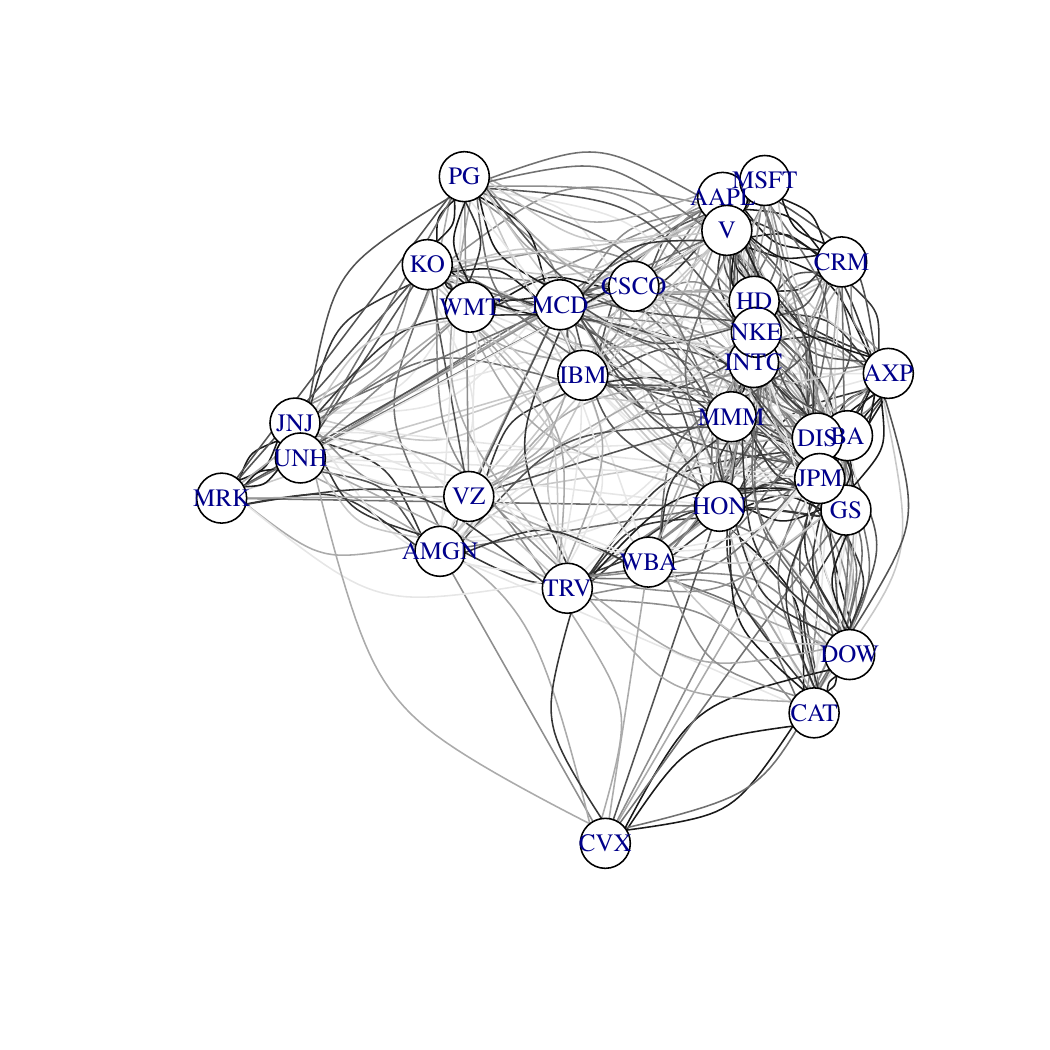}
     \includegraphics[width=0.48\textwidth]{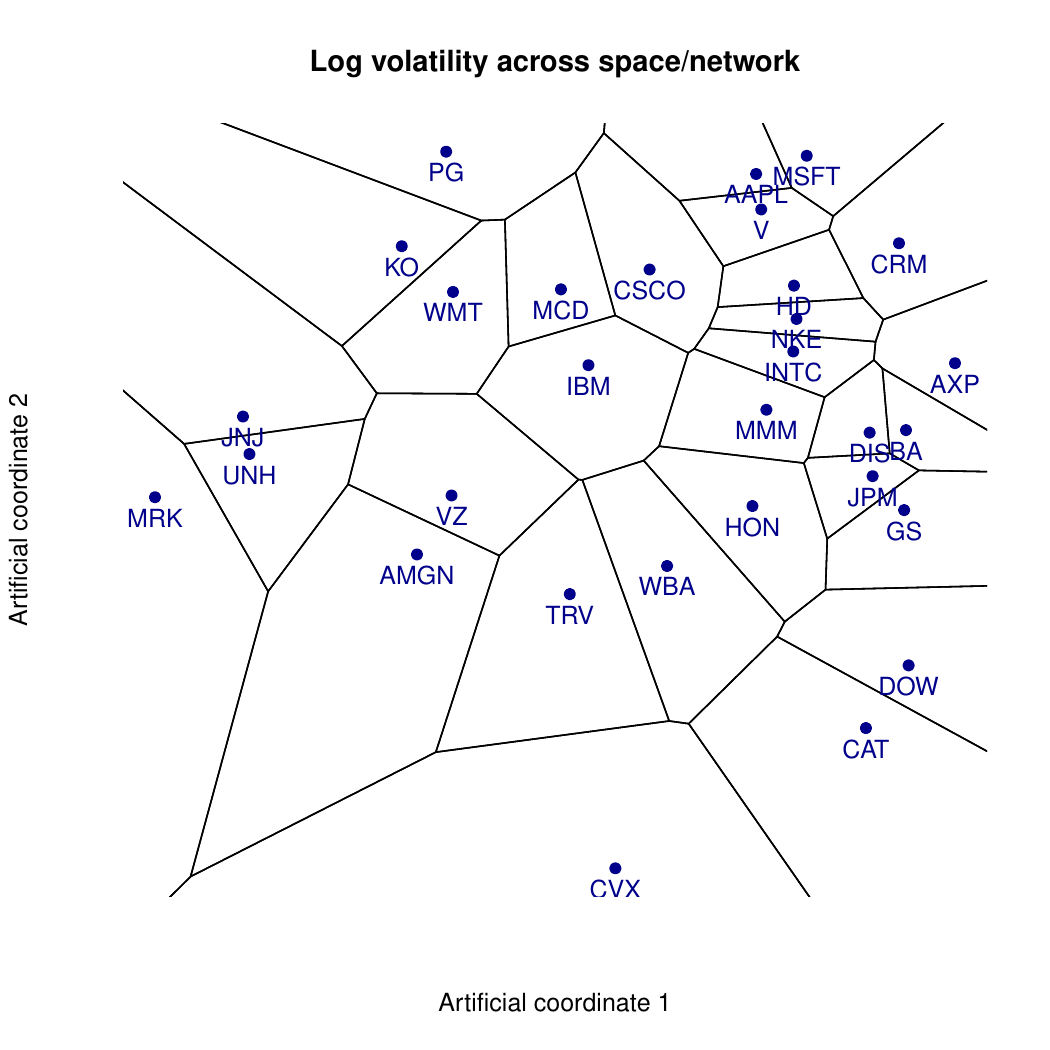}
	\end{center}
	\vspace{-0.80cm}
	\caption{Network and spatial representation of the $N=30$ stocks included in the Dow Jones Industrial Average Index. In the network representation (left), the nodes were placed such that highly correlated stocks appear close together, while weakly correlated stocks are positioned further apart. All edges with a weight greater than the median edge weight are displayed and colored according to their values, with darker colors representing higher weights. The network graph was additionally translated into a spatial representation (right) for visualization.}
	\label{fig:data_space}
\end{figure}

To construct the financial network matrix $\mathbf{M}$, we follow a consistent strand of literature \citep{onnela2004clustering,tumminello2010correlation,kumar2012correlation} that adopts a correlation-based distance to specify the elements of the network matrix. More specifically, we first define the correlation-based distance as
\begin{equation}
    d_{ij} = \sqrt{2(1-\rho_{ij})},
\end{equation}
where $\rho_{ij}$ is the Pearson correlation coefficient between two stocks $i$ and $j$. Given this transformation, the individual correlation coefficients are mapped from $[-1,1]$ to $[2,0]$, creating a symmetric distance matrix. We then define $m_{ij}=d^{-1}_{ij}$, so that a higher pairwise correlation coefficient indicates a stronger connection between the two stocks $i$ and $j$. Figure~\ref{fig:data_space} shows the network and spatial representation of the $N=30$ stocks included in the Dow Jones Industrial Average Index. In particular, highly correlated stocks appear close together in the network depicted in Figure~\ref{fig:data_space}, while weakly correlated stocks are positioned further apart. All edges with a weight greater than the median edge weight are displayed and colored according to their values, with darker colors representing higher weights.

Figure~\ref{fig:data_space} illustrates the existence of core-periphery patterns in the Dow Jones Industrial Average index. For instance, CVX is positioned farthest from the main cluster of stocks and is correlated with other peripheral stocks like DOW and CAT. Another group of peripheral stocks includes MRK, UNH, and JNJ, while IBM stands out as the most central stock. This is not surprising, as IBM is one of the oldest stocks and one of the largest in terms of market capitalization. The darkest links can be found in the right part of the network, which also represents its central region. This suggests that the stocks in the core of the network are the most correlated, while those in the periphery are less correlated with the core stocks. These results are particularly significant for volatility modeling, as the information from peripheral stocks can enhance the modeling of their volatility, while core information can be utilized to model the volatility dynamics of stocks located in the core of the network.

To apply the network log-ARCH model with common factors, it is essential to choose the number of factors, $q$. We estimate various models using Algorithm 1, varying the number of factors, and select the $q$ value that minimizes the DIC criterion. The DIC results for different $q$ values are presented in Table~\ref{tab:dic_stocks}. We compare the results obtained using Algorithm~\ref{a1} (standard MCMC) with those from the Bayesian Lasso approach (Algorithm~\ref{a2}, shrinkage MCMC), which automatically selects the relevant factors.

\begin{table}[!htb]
\centering
\begin{tabular}{lc}
\toprule
 & \textbf{DIC} \\
\midrule
\multicolumn{2}{l}{Algorithm \ref{a1}}\\
\midrule
$q=1$ & 9161.17 \\
$q=2$ & 9153.37 \\
$q=3$ & 9119.95 \\
$q=4$ & 9129.38 \\
$q=5$ & 9112.77 \\
$q=6$ & 9122.20 \\
$q=7$ & \textbf{9110.11} \\
$q=8$ & 9116.32 \\
\midrule
\multicolumn{2}{l}{Algorithm \ref{a2}}\\
\midrule
$q_{\max}=8$ & 9117.03 \\
\bottomrule
\end{tabular}
\caption{DIC values for Network log-ARCH models with a different number of common factors $q$ -- stock market data. Both Algorithm \ref{a1} and Algorithm \ref{a2} are considered. The best model, highlighted in bold, is the one with the lowest DIC value.}
\label{tab:dic_stocks}
\end{table}


In Table~\ref{tab:dic_stocks}, we observe that the Bayesian Lasso approach provides a larger DIC compared to Algorithm \ref{a1}. Therefore, following the results in Table \ref{tab:dic_stocks}, we choose the Network log-ARCH model with $q=7$ common factors estimated using Algorithm \ref{a1}. The selection of a relatively large number of latent factors is not surprising when analyzing volatility dynamics in the stock market. For instance, \cite{mittnik2015stock} identified five distinct categories of volatility factors: equity market, interest rates and bond market, exchange rates, liquidity, and macroeconomic variables. To these categories, we could also consider additional ones, for example, those based on behavioral variables \citep[see, e.g.,][]{akin2024behavioral}. The selection of $q=7$, therefore, appears consistent with the findings of previous studies.




The parameters $\beta$, $\gamma$, $\rho$, and $\delta$ correspond to the observable factor (market volatility), temporal effect, spatial effect (contemporaneous volatility network spillover), and spatiotemporal effect (lagged volatility spillover), respectively. The trace plots indicate stable behavior, suggesting that the MCMC chain produced by the algorithm converged. The estimated posterior medians for these parameters serve as our point estimates, and the results, along with their 95\% credible intervals, are presented in Table \ref{tab:estimates_stock}. The table displays the estimates obtained using Algorithm 1 with different numbers of factors $q$ and Bayesian LASSO. Statistically significant coefficients are those with the confidence interval that does not include zero.

\begin{table}[!htb]
    \centering
   \scalebox{0.8}{\begin{tabular}{lcccc|cc}
        \toprule
Factors   & $\beta$ & $\gamma$ & $\delta$ & $\rho$ & $\boldsymbol{\lambda}'\boldsymbol{f}_t$ & $\log h_t(\mathbf{s}_i)$ \\
        \midrule
        \multicolumn{7}{l}{Panel A: Algorithm \ref{a1}}\\
        
  $q=1$ & -2.3194 & -0.0163 & 0.2439 & 0.6014 & -0.0348 & -6.7389 \\
         & (-4.5869, -0.0627) & (-0.0426, 0.0098) & (0.1995, 0.2900) & (0.5596, 0.6400) & (-0.3224, 0.0000) & (-8.0607, -5.2877) \\
  $q=2$  & -2.5132 & -0.0090 & 0.2471 & 0.5980 & -0.0385 & -6.8176 \\
         & (-4.8519, -0.1461) & (-0.0379, 0.0197) & (0.1980, 0.2967) & (0.5546, 0.6392) & (-0.2582, 0.0679) & (-8.0593, -5.3428) \\
  $q=3$ & -2.4023 & -0.0033 & 0.2641 & 0.5810 & -0.0502 & -6.8203 \\
         & (-4.8424, 0.0232) & (-0.0327, 0.0253) & (0.1941, 0.3409) & (0.5077, 0.6462) & (-0.2859, 0.1142) & (-8.0861, -5.4150) \\
  $q=4$  & -2.2992 & 0.0007 & 0.2558 & 0.5906 & -0.0557 & -6.8724 \\
         & (-4.8152, 0.2018) & (-0.0298, 0.0308) & (0.1966, 0.3368) & (0.5108, 0.6446) & (-0.2608, 0.1030) & (-8.1370, -5.4373) \\
  $q=5$  & -2.2455 & 0.0007 & 0.2631 & 0.5868 & -0.0558 & -6.9205 \\
         & (-4.8613, 0.3050) & (-0.0308, 0.0319) & (0.2046, 0.3467) & (0.5020, 0.6388) & (-0.2723, 0.0899) & (-8.1628, -5.4741) \\
  $q=6$  & -2.2041 & 0.0007 & 0.2632 & 0.5873 & -0.0618 & -6.9161 \\
         & (-4.8054, 0.4158) & (-0.0309, 0.0325) & (0.2017, 0.3268) & (0.5298, 0.6446) & (-0.2794, 0.1057) & (-8.1435, -5.4985) \\
\rowcolor{lightgray}
  $q=7$  & -2.1727 & 0.0026 & 0.2695 & 0.5829 & -0.0923 & -6.9445 \\
\rowcolor{lightgray}
         & (-4.8516, 0.5328) & (-0.0286, 0.0336) & (0.2124, 0.3348) & (0.5155, 0.6353) & (-0.3164, 0.1536) & (-8.1203, -5.5319) \\
  $q=8$  & -2.1963 & 0.0006 & 0.2650 & 0.5911 & -0.1031 & -6.9823 \\
         & (-4.9415, 0.5523) & (-0.0316, 0.0327) & (0.2068, 0.3359) & (0.5173, 0.6395) & (-0.3033, 0.1476) & (-8.1459, -5.5332) \\
         \midrule
 \multicolumn{7}{l}{Panel B: Algorithm \ref{a2}}\\
    $q_{\max}=8$ & -2.3313 & 0.0112 & 0.2588 & 0.5892 & -0.0618 & -6.9758 \\
         & (-4.4242, -0.2384) & (-0.0265, 0.0489) & (0.2060, 0.3116) & (0.5472, 0.6313) & (-0.1867, 0.0630) & (-7.5128, -6.4389) \\
        \bottomrule
    \end{tabular}}
    \caption{Posterior median estimates with lower and upper bounds $(0.025, 0.975)$ of the Network log-ARCH with common factors -- stock market data. Different numbers of factors $q$ are in the rows, and the parameters are in the columns. The last row shows the results obtained from the Bayesian Lasso approach, using Algorithm~\ref{a2} with $q_{\max}=8$. Statistically significant coefficients are those with a confidence interval that does not include zero.}
    \label{tab:estimates_stock}
\end{table}


In Table~\ref{tab:dic_stocks}, we present the estimation results for $q$ values ranging from $1$ to $8$, with a focus on the results for $q=7$, which is highlighted in gray as the best model. In this case, both the parameters associated with the observable market volatility factor ($\beta$) and the own-nodes temporal lag ($\gamma$) are not statistically significant. However, we find that the instantaneous volatility spillover effect parameter ($\rho$) and the lagged cross-nodes volatility spillover effect parameter ($\delta$) are positive and statistically significant. This is notable because only the network information, through spatial and spatiotemporal effects, remains important once we account for the presence of latent factors. This also suggests that the past behavior of the entire network, adjusted for contemporaneous effects, is more influential than the lagged log return of the individual node (own past). Furthermore, the insignificance of the market volatility proxy suggests that this information is captured by the latent factors estimated in the model. Additionally, both the instantaneous and lagged spillover effects are positive, indicating that an increase in the volatility of adjacent nodes in the network is associated with a rise in volatility. The magnitude of the instantaneous effect $(0.58)$ is also larger than that of the lagged effect $(0.26)$. Also, note that the estimates obtained from the Bayesian Lasso approach using Algorithm~\ref{a2} with $q_{\max}=8$ are similar to those shown in the gray highlighted row.

The results obtained with other models are qualitatively very close to those obtained using $q=7$ latent factors. However, we think it is worth highlighting that in the case of $q=1$ and $q=2$, we find the observable market factor to be statistically significant. This suggests that up to two factors are not enough to remove the effect of market volatility or, said differently, none of the estimated two factors contain information about market volatility. All the estimated models, starting from $q=3$, instead show no significant $\beta$ coefficient. Considering also the DIC values shown in Table~\ref{tab:dic_stocks}, we find that these two simpler models with $q=1$ and $q=2$ are those with the lowest performance.

\begin{figure}[!htb]
	\begin{center}
	    \includegraphics[width=0.58\textwidth,trim = 0cm 0cm 0cm 0.6cm, clip = true]{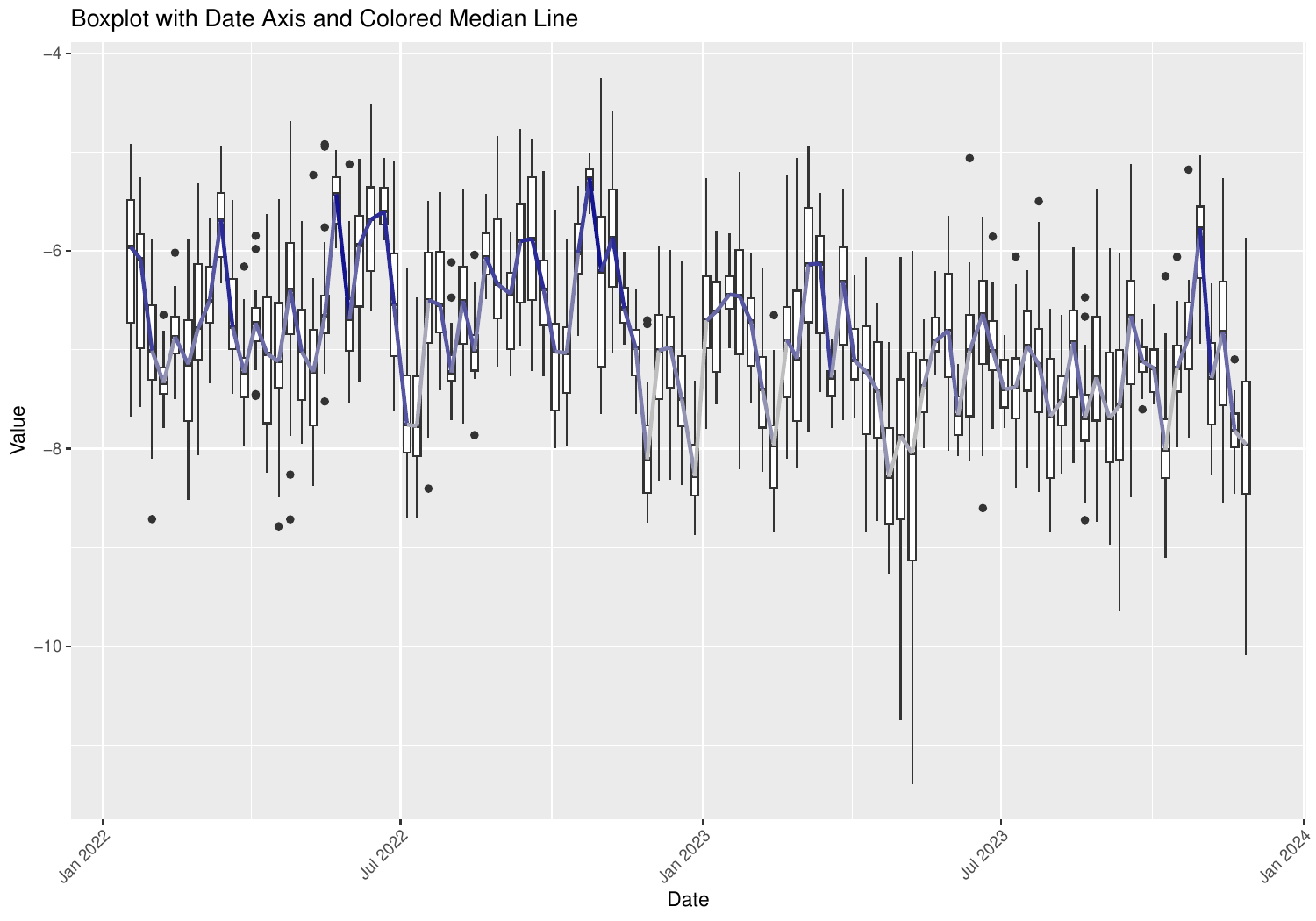}
     \includegraphics[width=0.41\textwidth]{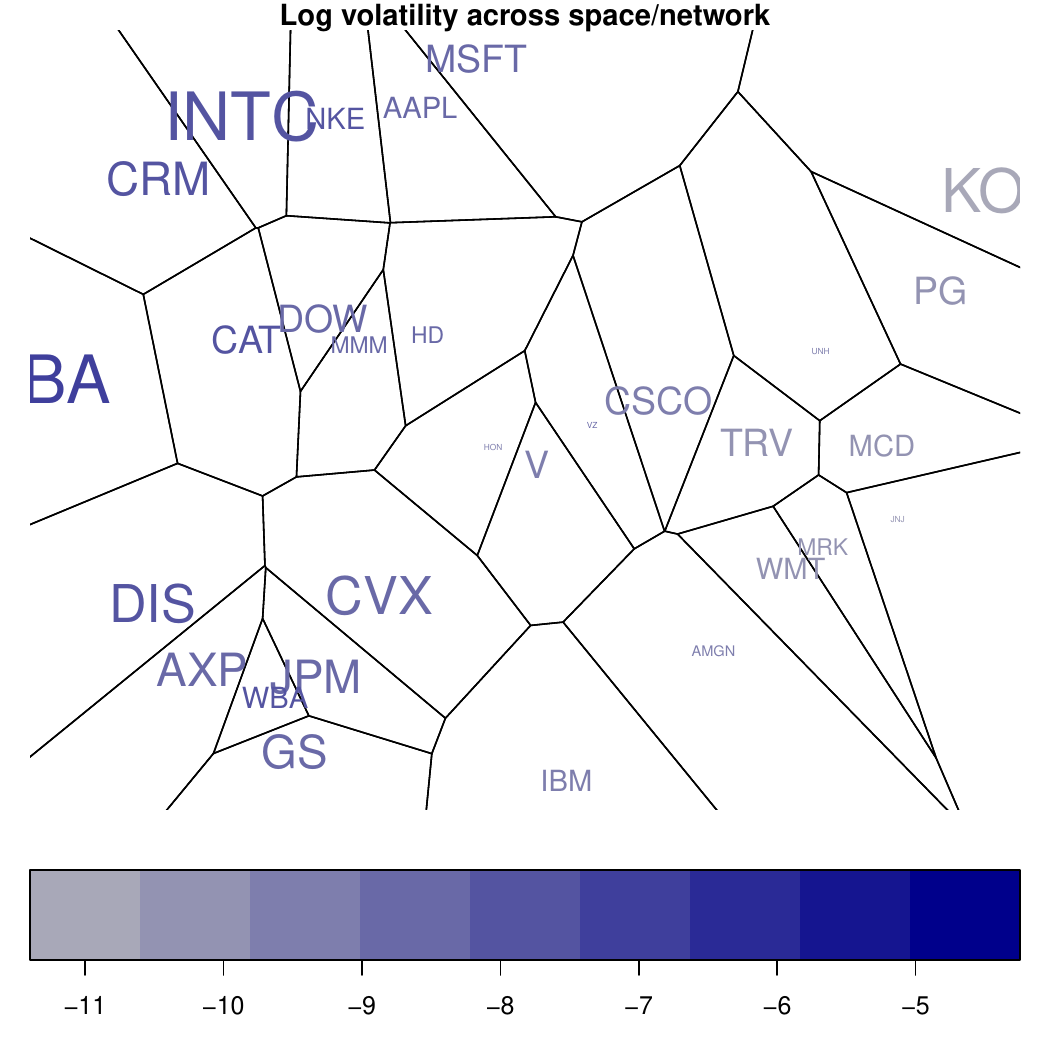}
	\end{center}
	\vspace{-0.80cm}
	\caption{Median posterior draws of the log volatilities $\widehat{\log h_{t}(\mf{s}_i)}$ plotted across time (left) and space/network (right) for 7 factors. For the time series plot, the estimated log volatilities are aggregated in the form of boxplots, and the median volatility is displayed as a solid central line. On the right, the spatial representation is shown analogously to Figure \ref{fig:data_space}. Stocks with similar volatility dynamics are placed in close proximity, while stocks with less correlated volatility dynamics are plotted further apart. The color of the labels is chosen according to the median log volatility across time, whereby the color scale of the left and right plots matches and can be used as a reference to relate both graphics. The larger the label size of the stock, the larger its interquartile range of the log volatilities, i.e., the risk/volatility profile is changing more over time compared to all other stocks.}
	\label{fig:estAlgo1}
\end{figure}

Finally, we show the estimated value of volatilities and the resulting network structure, highlighting similarities in the financial network. Figure \ref{fig:estAlgo1} shows the median posterior draws of the log volatilities obtained with Algorithm 1, assuming $q=7$ factors. The volatilities are plotted across time (left) and space/network (right). For the time series plot, the estimated log volatilities are aggregated in the form of boxplots, with the median volatility displayed as a solid central line. The interquartile ranges of the boxplots highlight a large variability in the volatility patterns for the stocks included in the DJIA index. For some periods, outlier stocks with exceptionally high (or low) volatility can be identified.

\FloatBarrier

On the right of Figure~\ref{fig:estAlgo1}, we display the estimated volatility patterns of the considered US stocks. Stocks with similar volatility dynamics are placed in close proximity, while stocks with less correlated volatility dynamics are plotted further apart. The color of the labels is chosen according to the median log volatility across time, whereby the color scale of the left and right plots matches and can be used as a reference to relate both graphics. The larger the label size of the stock, the larger its interquartile range of the log volatilities, i.e., the risk/volatility profile is changing more over time compared to all other stocks. The figure highlights the existence of two groups of stocks: those with high and low volatility levels. In particular, BA, CRM, and INTC are the stocks with the highest volatility values and are those closer in the network. The same applies to AXP, DIS, and CVX, albeit with a lower level of volatility compared to the previous ones. On the other side of the network, we find many stocks with lower levels of variability, namely AMGN, WMT, and MRK. HON and VZ, the stocks with the lowest volatility, are neighbors and located in the center of the network structure.

In summary, Figure \ref{fig:estAlgo1} confirms the existence of positive volatility spillovers because stocks with large volatility are closer to those with similarly large volatility levels, and the same applies to stocks with lower variability. This evidence corroborates the findings of previous studies on the existence of relevant spillovers in the stock market and on the usefulness of using network methods for modeling financial volatility.


\section{Conclusion}\label{sec6}

This paper introduced a dynamic spatiotemporal and network log-ARCH model that incorporates spatial, temporal, and spatiotemporal spillover effects with common factors. A key innovation of this model is the inclusion of volatility-specific time-varying latent factors that evolve over time, capturing complex volatility dynamics more effectively. Unlike conventional methods, this approach integrates both spatial dependencies—how volatility in one location affects another—and temporal evolution, providing a richer, more detailed understanding of volatility patterns across time and space. The model also has a network interpretation, linking different regions or assets through their volatility dependencies, further enriching the understanding of how shocks propagate across space and time.

We suggest two Gibbs samplers to estimate the parameters of the model. In addition to parameter estimates, our Bayesian estimation approach also allows us to estimate the volatility of the outcome variable. We suggest using the DIC criterion and a Bayesian Lasso approach to choose the number of factors that yield better predictive performance.  We provide simulation results demonstrating that our proposed algorithms perform satisfactorily in recovering the true parameter values and volatility.

The proposed model is particularly well-suited for analyzing markets where spatial dependencies play a crucial role, such as the US housing market, but it can also be applied for modeling more general types of network processes. Two practical applications are proposed: one analyzing house price volatility and the other focusing on stock market data. In both cases, the model's ability to capture spatial and spatiotemporal dynamics offers a more comprehensive view of volatility transmission mechanisms, highlighting the model's broad applicability and its potential to enhance decision-making in financial and real estate markets.

In the first application, the US housing market provides a natural context for applying spatiotemporal models, given the established evidence of spatial spillovers in house prices. The second application extends this model to stock market data, where volatility spillovers are known to occur across different stocks and market indices. The empirical findings reveal that positive spillovers are present in both the housing and stock markets. However, network spillovers are much more pronounced in the stock market, where assets are more tightly interconnected, and volatility spreads more rapidly across the network. These findings highlight the model's ability to capture varying degrees of interconnectedness across different markets, making it a versatile tool for understanding complex volatility dynamics.

While the proposed approach is useful for volatility modeling, less is known about its forecasting performance. This is particularly interesting in both empirical applications we have considered in the paper. On one hand, in the case of house prices, forecasting volatility is an important indicator of economic stability, and cycles in housing market volatility can be used for predicting crises. On the other hand, in the case of financial markets, forecasting volatility is a useful tool for portfolio selection. Moreover, accurate volatility forecasts can also be used by policymakers to track conditions in the stock market and may help prevent crises. Therefore, a future study on the forecasting performance of our model, comparing its performance with other suggested models in the literature, can be explored. Moreover, future research may extend the proposed method to accommodate more complex factorial structures, such as the existence of a cluster-level structure.

\end{document}